\documentclass[usenatbib]{mnras}
\usepackage{amsmath}
\usepackage{amssymb}
\usepackage{morefloats}
\usepackage{graphicx}
\usepackage{array}
\usepackage{siunitx}
\usepackage{subfig}

\newcolumntype{L}[1]{>{\raggedright\let\newline\\\arraybackslash\hspace{0pt}}p{#1}}
\newcolumntype{C}[1]{>{\centering\let\newline\\\arraybackslash\hspace{0pt}}p{#1}}
\newcolumntype{R}[1]{>{\raggedleft\let\newline\\\arraybackslash\hspace{0pt}}p{#1}}

\title[A Bayesian approach to dust in CCSNe]{Measuring dust in core-collapse supernovae with a Bayesian
	approach to line profile modelling}

\author[Antonia Bevan]{Antonia Bevan$^{1}$\\
$^{1}$Department of Physics and Astronomy, University College London, 
Gower Street, London WC1E 6BT, UK}

\begin{document}

\date{Accepted on 31 July 2018}

\pagerange{\pageref{firstpage}--\pageref{lastpage}} \pubyear{2018}

\maketitle

\label{firstpage}

\begin{abstract}

Optical and near-IR (NIR) line profiles of many ageing core-collapse supernovae (CCSNe) exhibit an apparently asymmetric bluewards shift often attributed to greater extinction by internal dust of redshifted radiation emitted from the receding regions of the SN ejecta. The {\sc damocles} Monte Carlo line radiative transfer code models the extent and shape of these dust-affected line profiles to determine the dust mass that has condensed, in addition to other properties of the dusty ejecta. I present here the application of an affine invariant Markov Chain Monte Carlo (MCMC) ensemble sampler (emcee) to the {\sc damocles} code in order to investigate the multi-dimensional parameter space rigorously and characterise the posterior probability distribution. A likelihood function is formulated that handles both Monte Carlo and observational uncertainties.  This Bayesian approach is applied to four simulated line profiles in order to test the method and investigate its efficacy.  The majority of parameters can be tightly constrained using this method, and a strong (predictable) dependence between the grain size and the dust mass is quantified. The new approach is also applied to the H$\alpha$ line and [O {\sc i}]\,6300,\,6363\,\AA\ doublet of SN~1987A at 714\,d post-outburst, re-examining a previous 5-dimensional smooth model and also investigating a new, more complex, 10-dimensional model that treats both features simultaneously.  The dust mass, dust grain size and a range of other parameters can be well constrained using this technique, representing a significant improvement over the previous manual approach.

\

\end{abstract}

\begin{keywords}
	line: profiles --
	radiative transfer --
	methods: statistical --
	supernovae: general --
	supernovae: individual: SN 1987A  --  
	ISM: supernova remnants.
\end{keywords}

\section{Introduction}

There remain numerous questions surrounding the formation of dust in the universe. Significant challenges are still faced in the determination of dust formation rates, mechanisms and environments.  Motivated by seemingly inexplicably large masses of dust observed at high redshifts \citep{Omont2001,Bertoldi2003,Watson2015,Laporte2017}, there is a widespread 
desire to understand the nature of the primary sources of dust in the universe.  

Core-collapse supernovae (CCSNe) are known to produce dust in their ageing ejecta. Theoretical models  predict that CCSNe are capable of producing $>0.1$M$_{\odot}$ of ejecta-condensed dust \citep{Todini2001,Nozawa2003,Gall2011,Sarangi2015}. To date, however, dust masses in the majority of these objects have been inferred via fits to their near-infrared (NIR) and mid-infrared (MIR) spectral energy distributions (SEDs), which trace only warm dust.  Warm dust masses up to $\sim10^{-3}$~M$_{\odot}$ have been detected at late times ($>1$~yr) in several CCSNe \citep{Sugerman2006,Meikle2007,Andrews2010,Fabbri2011,Gall2011,Gomez2013,Gall2014}.  However, a few objects have also been observed in the far-IR, allowing their full SEDs to be fitted and therefore tracing the presence of cold dust as well as warm and hot dust.  Using this technique, dust masses $\gtrsim0.1$\,M$_{\odot}$ have been estimated to have formed in SN~1987A, Cassiopeia~A and the Crab Nebula \citep{Gomez2012, Indebetouw2014,Matsuura2015,deLooze2017,Owen2015}. More recently, it has been suggested that a significant mass of dust (0.08~--~0.9~M$_{\odot}$) has also formed in the Galactic supernova remnant G54.1+0.3 \citep{Temim2017,Rho2017} as well as very large masses of dust ($>1$\,M$_{\odot}$) in a number of other Galactic supernova remnants (Chawner et al. in prep.). An average net dust production rate of 0.1~--~1.0~M$_{\odot}$ per CCSN is required to account for the dust masses observed in the early universe \citep{Morgan2003,Dwek2007}. 
The dust budget problem would therefore be resolved if 
these few objects were, in fact, representative of the wider CCSN population and the dust was able to survive the passage of the reverse shock \citep{Bianchi2007,Bocchio2016}. A larger sample of CCSN dust mass estimates is therefore required.

Following the end of the \textit{Herschel} mission in 2013, there will be a long wait for instruments that are capable of detecting cold dust emission at far-IR wavelengths and so an alternative approach is needed.  The {\sc damocles} Monte Carlo line radiative transfer code predicts dust masses in the ejecta of CCSNe by modelling the red-blue asymmetry frequently observed in their optical and NIR emission lines \citep[hereafter B16]{Bevan2016}.  This asymmetry is due to the condensation of dust in the ejecta causing redshifted radiation from the receding regions of the supernova to experience greater extinction than blueshifted radiation emitted from the approaching regions \citep{Lucy1989}. In addition to providing an alternative method for tracing both warm and cold dust in the ejecta, line profile modelling has the added advantage of tracing only newly-condensed dust within the ejecta. Pre-existing circumstellar dust may  contribute to the observed flux in the IR but the red and blue components of optical or NIR lines emitted from within the ejecta will be similarly attenuated by the surrounding circumstellar dust, i.e. any dust-induced red-blue asymmetry must be solely a result of internal, ejecta-condensed dust. 

The approach also allows other properties of the dust to be determined. Of particular interest is the dust grain size distribution.   Regardless of the masses of dust that form in the ejecta of CCSNe, the grains will eventually be subject to a reverse shock that will pass back through the ejecta, potentially destroying these newly-formed grains and significantly diminishing the dust mass that has formed.  The size of dust grains that condense within the supernova ejecta  determine their likelihood of survival. An understanding of dust grain sizes in CCSNe is therefore critical to determining the relative contribution of CCSNe to dust production in the universe.

B16 applied the {\sc damocles} Monte Carlo code to the H$\alpha$ and [O{\sc i}]\,6300,6363\,\AA\ lines of SN~1987A between 714\,d and 3500\,d post-explosion. A steady increase in the ejecta dust mass over this period was inferred with a predicted current dust mass of 0.8~M$_{\odot}$, consistent with dust mass estimates derived from SED fitting and modelling \citep{Matsuura2011,Indebetouw2014,Matsuura2015,Wesson2015}.  The {\sc damocles} code was also applied to the late-time optical line profiles of SN~1980K, SN~1993J and Cas~A by \citet{Bevan2017}.  Dust masses of 0.12~-–~0.3\,M$_{\odot}$ at 30~yr, 0.08~–-~0.18\,M$_{\odot}$ at 16~yr and $\sim$1.1\,M$_{\odot}$ at $\sim$330~yr were predicted respectively \citep{Bevan2017}. Clearly, further examples are needed in order to establish whether this apparent trend towards larger ($\gtrsim0.1$~M$_{\odot}$) dust masses is an accurate representation of dust formation in CCSNe more generally. 

In working towards the overall goal of understanding the masses and properties of dust in the ejecta of CCSNe, I have explored the implementation of a Bayesian methodology for line profile fitting. The fundamental power of Bayesian statistics in providing a framework to understand the probability of a model when the the data is known has been increasingly exploited in astronomy over the last twenty years \citep[e.g.][]{Strolger2004, Venn2004,Ilbert2006, Feroz2009, Arzoumanian2016}. In particular, Markov Chain Monte Carlo (MCMC) methods have provided efficient, robust and rigorous procedures with which to explore highly multi-dimensional parameter spaces and quantify posterior probability distributions.  Increasingly available computing power has allowed these methods to be employed in a wide variety of fields with impressive results that yield significantly more insight than can be gained from a single best-fitting set of parameters. \citet{Sharma2017} presents a comprehensive review of MCMC methods for Bayesian data analysis in astronomy.  

 I have applied an affine invariant ensemble sampler \citep{Goodman2010} to the {\sc damocles} code in order to map the multi-dimensional posterior probability distribution of a range of models and parameter spaces. I initially employed the sampler to model four simulated, or `theoretical', line profiles (generated by {\sc damocles}) that were deemed representative of observed line profiles of CCSNe at late times but for which the `true solution' was known (models A1 - A4).  
I also revisited the models by B16 of the H$\alpha$ line and the [O~{\sc i}]\,6300,6363\,\AA\ doublet of SN~1987A at 714\,d.  The new approach was applied to a smooth, 5-dimensional model based on their work (model B). A new model (model C) was also explored which treated both emission features simultaneously and used the newly-applied Bayesian methodology to characterise a significantly more complex 10-dimensional  parameter space.  
The ultimate goal is to assess the validity of this approach with regard to its application to both archival and future datasets, with a view to significantly expanding the current range of dust mass estimates for CCSNe years after outburst.  

In Section \ref{the_problem}, the formulation of the problem is presented along with a discussion of how Monte Carlo and observational uncertainties are handled and a brief description of the affine invariant ensemble sampler.  The adopted priors, variable parameters and posterior distributions for all models are presented in Section \ref{scn_results}.  I discuss the implications of these results in Section \ref{scn_discussion} and compare them to results obtained from previous manual line profile fitting using {\sc damocles}, as well as results obtained from SED fitting.  The constraints placed on other parameters such as dust grain size and dust density distribution are also discussed. I emphasise the potential for future application to other objects before summarising and concluding in Section \ref{conclusions}.


{\renewcommand{\arraystretch}{1.5}%
	\begin{table*}
		\centering
		\caption{Values of the parameters adopted to generate four representative simulated line profiles (A1 - A4). In these models, amorphous carbon grains were used with the optical constants of \citet{Zubko1996}. The H$\alpha$ line profile was modelled with an intrinsic smooth radial power-law emissivity distribution ($i \sim r^{-2\beta}$) applied to a homologously expanding shell geometry at day 1000.}

		\begin{tabular*}{\linewidth}{L{1cm} C{3.4cm} C{2.2cm} C{2.2cm} C{1.9cm} C{2.2cm}  c}
			\hline
			& &$v_{\rm max}$ & $v_{\rm min}$ & $\beta$ & $\log a$ &  $\log M_{\rm dust}$  \\ 
			& &10$^3$\,km\,s$^{-1}$ & 10$^3$\,km\,s$^{-1}$ && $\log \mu$m & $\log\,$M$_{\odot}$   
			\\ \hline
			A1 &``typical" &4.0          & 1.2          & 2.50     & -1.0  & -4.6                       \\ 
			
			A2 & ``double peaked" &4.0 &2.8 &1.0 &-1.0 & -4.6 \\
			A3 &``large grains"&4.0 &1.2 &2.0 & 0.18 & -3.9 \\
			A4 &``strongly blueshifted" &4.0 &1.2 &2.5 &-1.6 & -3.6 \\
			
			\hline
		\end{tabular*}
		
		\label{tb_synthetic_profiles}
		
	\end{table*}

\section{Formulation of the Bayesian approach}
\label{the_problem}

\subsection{\sc damocles}
{\sc damocles} is a Monte Carlo radiative transfer code that models the effects of dust, composed of any combination of species and grain size distributions, on optical and NIR emission lines emitted from the expanding ejecta of a late-time ($>1$\,yr) supernova.  For full details of the code and its testing, please see B16. By default, both the emissivity distribution and the dust distribution follow smooth  radial power-law distributions although any arbitrary distribution may be specified by providing the appropriate grid.  {\sc damocles} will also treat a variety of clumping structures as specified by a clumped dust mass fraction, volume filling factor, clump size and clump power-law distribution. The emissivity distribution may also initially be clumped.  The code has a large number of variable parameters ranging from 5 dimensions in the simplest, smooth models to $>20$ in the most complex cases.

\subsection{The Bayesian approach}

The aim is to map the posterior probability distribution based on the observations and our prior understanding of the physical situation.  The \textit{posterior} is defined by Bayes' Theorem as

\begin{equation}
\label{bayes_thm}
P(\boldsymbol \theta \,|\,\boldsymbol D) = P(\boldsymbol \theta ) \, \frac{P(\boldsymbol D\,|\,\boldsymbol \theta)}{P( \boldsymbol D)}
\end{equation}

\noindent where $\boldsymbol D$ represents the data that we wish to analyse (in our case, the observed or simulated line profile) and $ \boldsymbol \theta$ represents the parameters of our model. $P(\boldsymbol \theta )$ therefore represents our prior understanding of the probability of the model parameters (\textit{the prior}), $P(\boldsymbol D\,|\,\boldsymbol \theta)$ is the probability of obtaining the data for a given set of model parameters (\textit{the likelihood}) and $P(\textbf{D})$ is the probability of the data for all models (\textit{the evidence}). Since $P(\boldsymbol D)$ is independent of $\boldsymbol \theta$, we will only be interested in the scaled posterior as defined by 

\begin{equation}
\label{bayes_thm2}
P(\boldsymbol \theta\,|\,\boldsymbol D) \propto P(\boldsymbol \theta ) \, P(\boldsymbol D\,|\,\boldsymbol \theta)
\end{equation}

The posterior distribution will allow us to understand relationships between the parameters and to visualise which are the most likely regions of parameter space.  The aim is not to identify the single `best-fitting' model but to map the variation of likelihood across the entire space.   The prior is the probability before looking at any data.  It can be driven by theoretical models, previous observations or physical intuition. The likelihood is, practically, a mechanism for forward modelling i.e. simulating the data given a model and its parameters. It is proportional to $\exp({-\chi^2/2})$, where $\chi^2$ is the standard metric typically used to compare data and models in frequentist techniques.

In order to characterise the target posterior distribution, we may draw samples from across the parameter space. A single sample in parameter space is translated to a point in the target posterior distribution via Equation \ref{bayes_thm2}.  A likelihood function that describes the relationship between the model and the data must therefore be defined, and a prior probability distribution for each parameter must also be specified based on our current knowledge (e.g. physical constraints). Once defined, the ensemble sampler can be employed with this likelihood function in order to map the complete posterior distribution.

\subsection{Affine Invariant MCMC Ensemble Sampler}

There are numerous MCMC algorithms that work out how to sample points in parameter space efficiently in order to converge on a stable solution as quickly as possible.  In this work, I used the Python package `emcee' \citep{emcee}.  This package uses an affine invariant ensemble sampler as described by \citet{Goodman2010}, to which publication the reader is referred for full details of the algorithm. I present a summary below.

The ensemble sampler acts on a collection of points (n-dimensional position vectors in parameter space) termed `walkers'.  An initial position for each walker is sampled according to a distribution specified by the user.  The likelihood $P(\boldsymbol D\,|\,\boldsymbol \theta)$  of the model corresponding to the current set of parameters is calculated. A new point in parameter space is sampled based on the current positions of the other  walkers and the likelihood of this new point is also calculated. The ratio of these likelihoods determines whether the position of the walker is updated or not (i.e. the new point is either accepted or rejected). 
As such, the walkers `walk' around the entire parameter space exploring the posterior distribution in such a manner that the value of the posterior distribution in a given region of parameter space is characterised by the density of walkers in that region.    Faster convergence is attained when the walkers are initialised near regions of high likelihood but they will explore the entire space regardless of their initial positions. The extent of the space to be explored is determined by the bounds of the prior distributions (if applicable).

\begin{figure}
	\includegraphics[clip = true, trim = 20 10 80 20,width = \linewidth]{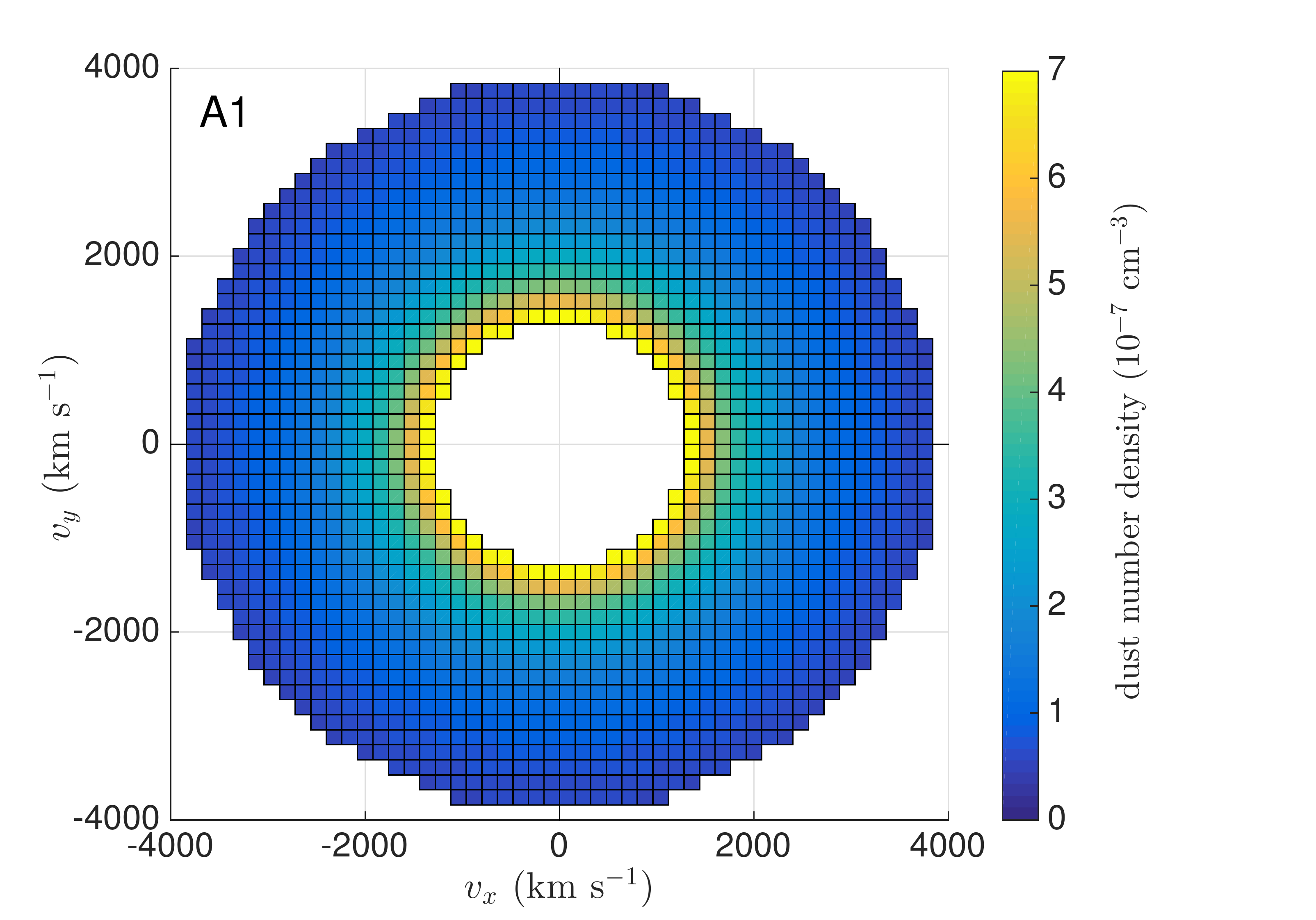}
	\includegraphics[clip = true, trim = 20 10 80 20,width = \linewidth]{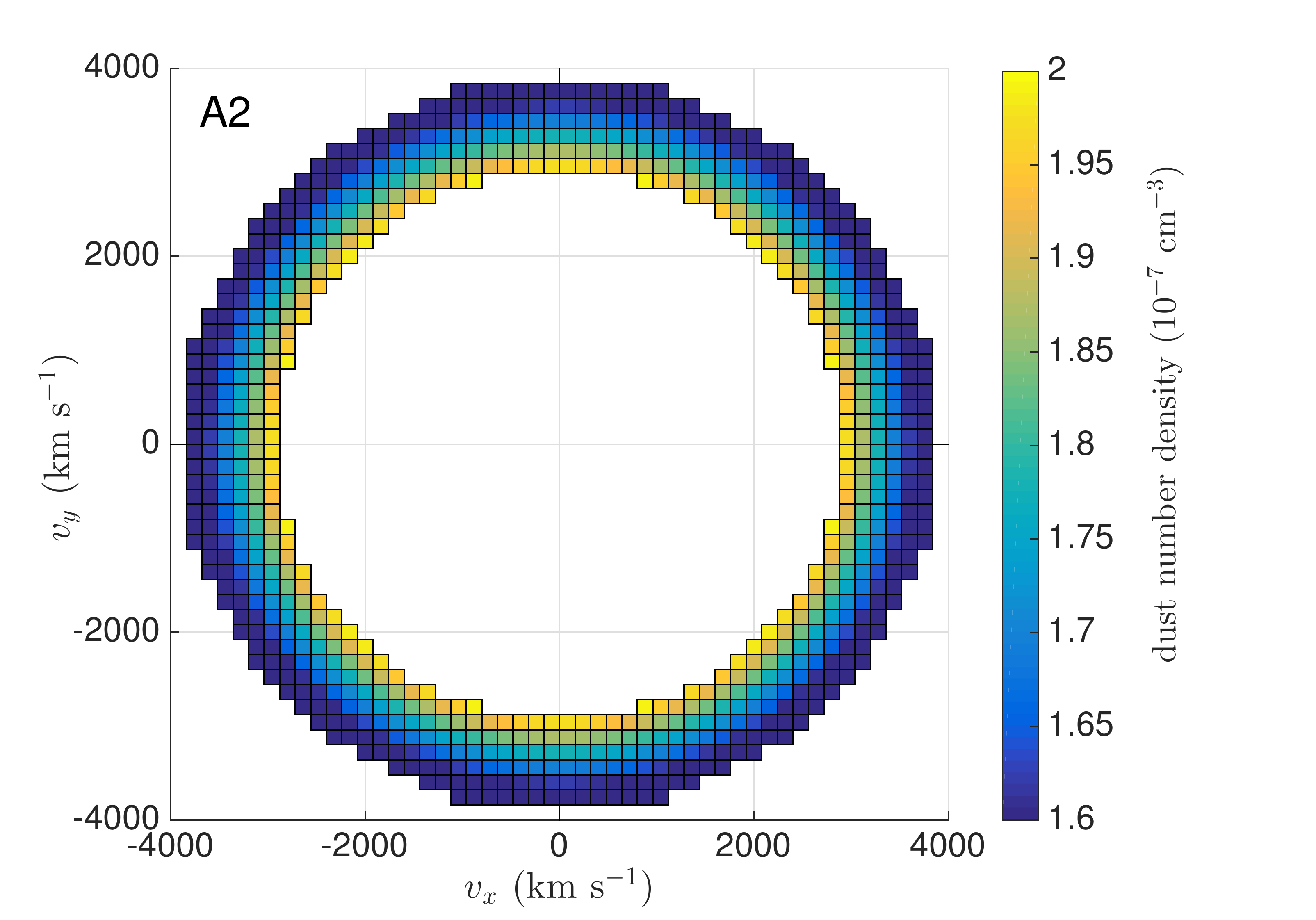}
	\caption{Schematics illustrating the shell geometries with smooth, radial dust density power-laws used for models A1 and A2.  \textit{Above:} Model A1 with $v_{\rm max} = 4000$\,km\,s$^{-1}$, $R_{\rm in}/R_{\rm out} = 0.3$ and $\rho \propto r^{-2.5}$. \textit{Below:} Model A2 with $v_{\rm max} = 4000$\,km\,s$^{-1}$, $R_{\rm in}/R_{\rm out} = 0.7$ and $\rho \propto r^{-1.0}$. The grids are divided into 50 in each axis.}
	\label{fig_shells}
\end{figure}

It is, of course, possible simply to grid parameter space, evaluate the likelihood and prior at each point and multiply them to get the posterior. However, whilst this would be exact, it would also be incredibly intensive and likely impossible for $>4$ dimensions. MCMC methods approximate the posterior by exploring the parameter space intelligently and are therefore a popular alternative. {\sc damocles} has between 5 and 20 variable parameters, strong degeneracies between certain of these parameters but no multimodality.  MCMC methods are extremely well-suited to this regime and are therefore an ideal choice.

The choice to use this particular MCMC methodology was made for a number of reasons.  Affine transformations are those which preserve the relative positions of points, lines and planes, for example reflection, rotation and scaling are all affine transformations.  This algorithm is designed to be affine invariant such that the parameter space can be `stretched' in order to sample points from a more isotropic distribution.  This ensures that it requires very little tuning in order to obtain good performance and is in contrast to a number of other MCMC algorithms such as Metropolis-Hastings \citep{Allison2014}. This property makes the algorithm particularly useful for models with parameters that range over significantly different scales as here. In addition to this, the ease of use and implementation (via the emcee package) and its speed and efficiency for problems with dimensionality of this order led to the choice of this algorithm over other available options.

	\begin{figure*}
	\subfloat{\includegraphics[clip = true, trim = 20 200 50 220, width = 0.5\linewidth]{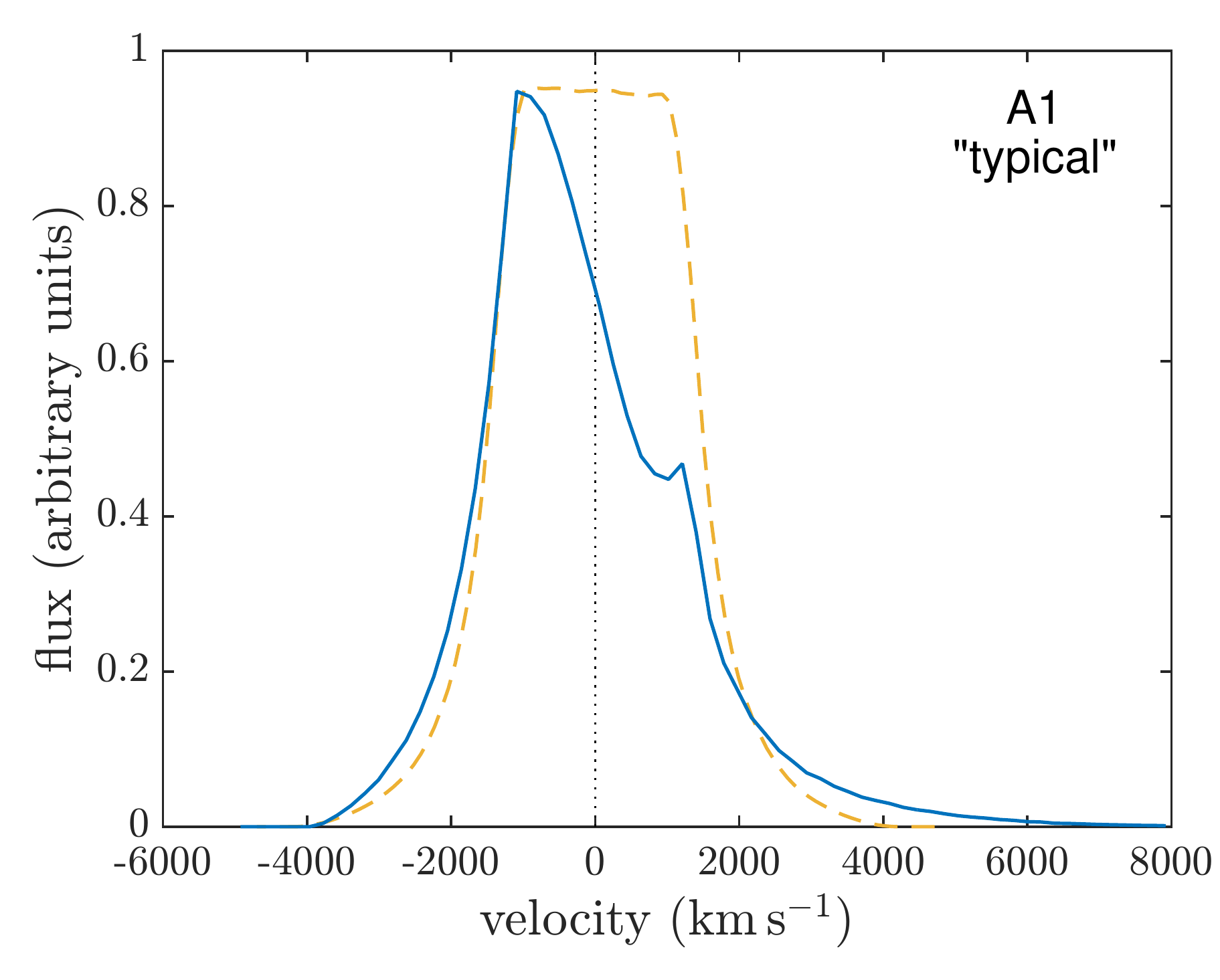}}
	\subfloat{\includegraphics[clip = true, trim = 20 200 50 220, width = 0.5\linewidth]{{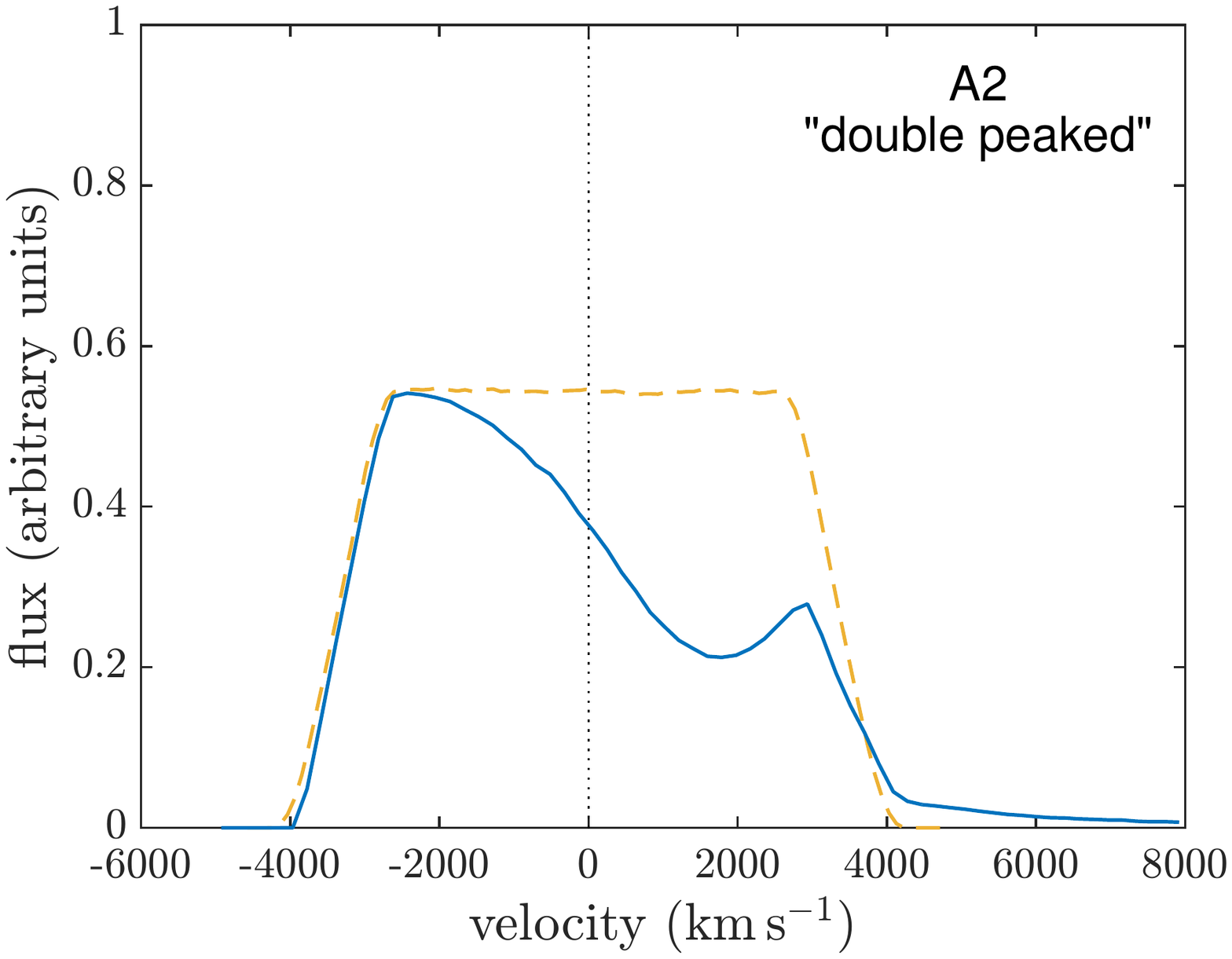}}}
	\\
	\subfloat{\includegraphics[clip = true, trim = 20 200 50 220, width = 0.5\linewidth]{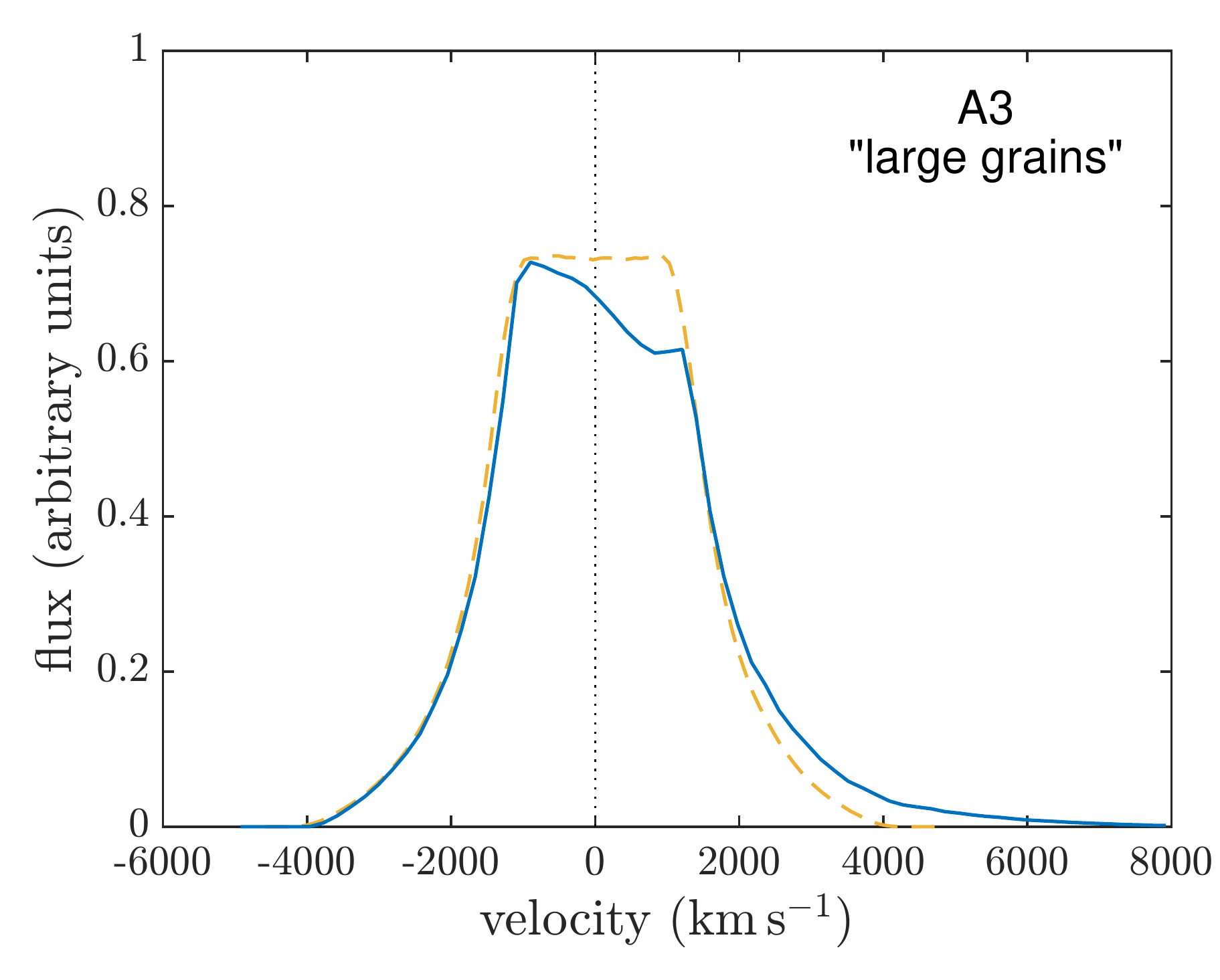}}
	\subfloat{\includegraphics[clip = true, trim = 20 200 50 220, width = 0.5\linewidth]{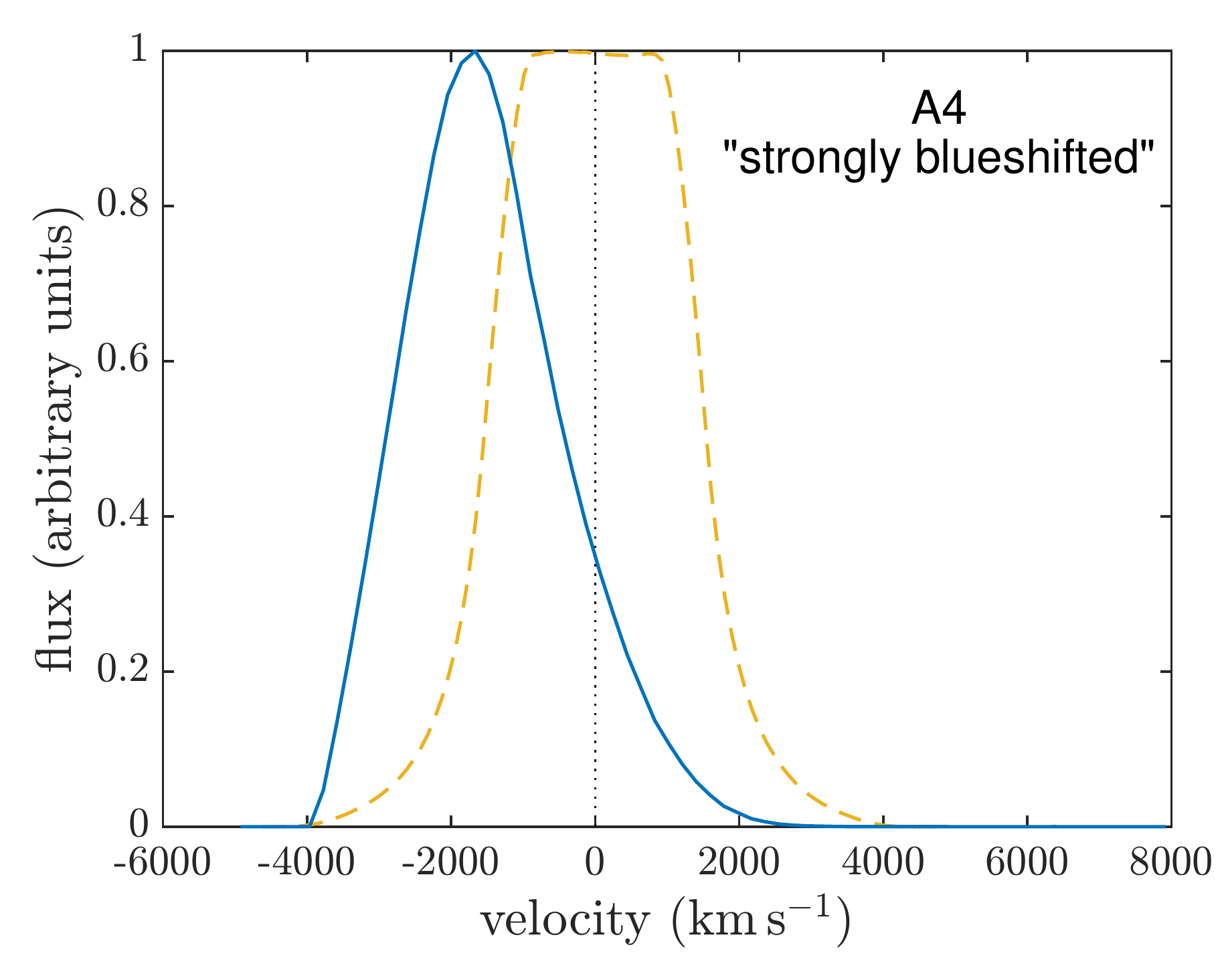}}
	\caption{\textit{Blue solid lines:} Four simulated line profiles generated using {\sc damocles} representing different types of dust-affected line profiles corresponding to the parameters listed in Table \ref{tb_synthetic_profiles}. \textit{Yellow dashed lines:} The corresponding intrinsic lines profiles with no dust present and scaled to the same peak flux.}
	\label{fig_sim_line_profiles}
\end{figure*}

\subsection{Formulating the likelihood function}

In order to quantify the likelihood of a particular set of model parameters given the observational or simulated data, we must define a function that relates the model to the data.  We here base the likelihood function on the typical $\chi^2$ comparison which is defined as 

\begin{equation}
\label{eqn_likelihood}
\chi^2 = \sum_{\rm i=1}^{n} \frac{\left( f_{\rm mod,i}-f_{\rm obs,i}\right) ^2}{\sigma_{\rm i}^2}
\end{equation}

\noindent  where $f_{\rm mod,i}$ and $f_{\rm obs,i}$ are the model flux and observed flux in frequency bin $i$ respectively.  $\sigma_i$ represents the overall uncertainty in frequency bin $i$ and $n$ is the number of frequency bins in the observed line profile.  

There are two primary contributions to the uncertainty $\sigma_i$ in each bin: there is an inherent uncertainty on the observational data and there is also uncertainty arising from the statistical nature of the Monte Carlo radiative transfer simulation.

{\renewcommand{\arraystretch}{1.5}%
	\begin{table*}
		\caption{The adopted prior distributions for the variable parameters in each model: the simulated line profile models A1 - A4, the smooth, 5-dimensional model of the SN~1987A H$\alpha$ line at 714\,d (model B), and the clumped, simultaneous H$\alpha$ and [O~{\sc i}]\,6300,\,6363\,\AA\ model of SN~1987A at 714\,d (model C). $U(a,b)$ indicates that the variable is distributed uniformly between $a$ and $b$.  $v_{\rm max}$ is the maximum velocity, $v_{\rm min}$ is the minimum velocity, $\beta_{\rm smooth}$ is the steepness of the smooth power-law density distribution, $\beta_{\rm clump}$ is the steepness of the power-law clump number density distribution, $M_{\rm d}$ is the dust mass, $f$ is the clump volume filling factor, $a$ is the dust grain radius, and $F_{\rm 6300}/F_{\rm 6363}$ is the flux ratio of the 6300\AA\ and 6363\AA\ components of the [O~{\sc i}]\,6300,\,6363\,\AA\ doublet.}
		\label{tb_priors}
		\centering
		\begin{tabular*}{\linewidth}{l  p{2.0cm}  C{2.9cm} C{2.9cm} C{2.9cm} C{1.7cm}}
			\hline
			\multicolumn{2}{l}{Parameter} & Simulated \newline line profiles \newline (A)                                                & SN 1987A \newline H$\alpha$ smooth \newline (B) & SN 1987A \newline H$\alpha$ and [O{\sc i}] clumped \newline simultaneous (C) &  Units  \\ \hline
			\multicolumn{2}{l}{Dust: }  &&&& \\
			& $v_{\rm max}$ 		& $U(3.0,8.0)$	& $U(2.0,5.0)$&$U(2.0,6.0)$&10$^3$\,km\,s$^{-1}$\\
			& $v_{\rm min}$ 		& $U(0.5,3.0)$  	& $U(0.1,1.1)$&$U(0.1,1.5)$& 10$^3$\,km\,s$^{-1}$\\ 
			& $\beta_{\rm smooth}$ 				& $U(1.0,3.0)$ 		& $U(0.2,2.0)$&-&\\
			&  $\beta_{\rm clump}$ & -&- &$U(0.0,3.5)$ & \\
			
			& $\log\,M_{\rm d}$ 	& $U(-6.0,-2.0)$		& $U(-7.0,-2.8)$&$U(-6,-3.5)$ &$\log\,M_{\odot}$\\ 
			& $f$					& - 				&-	&$U(0.1,0.7)$& \\
			& $\log\,a$ 	& $U(-2.0,0.7)$	& $U(-3.0,0.7)$	&$U(-2.0,0.7)$ &$\log\,\mu$m\\ \hline
			\multicolumn{2}{l}{H$\alpha$:} 	&&&& \\ 
			& $v_{\rm max}$ 		& coupled to dust 	&coupled to dust&coupled to dust&10$^3$\,km\,s$^{-1}$ \\ 
			& $v_{\rm min}$   		& coupled to dust 	&coupled to dust&$U(0.1,1.5)$&10$^3$\,km\,s$^{-1}$ \\ 
			& $\beta$ 			& coupled to dust 	&coupled to dust	&$U(0.0,2.0)$& \\ \hline
			\multicolumn{2}{l}{[O{\sc i}]:} \\ 
			& $v_{\rm max}$ 		 		& - 				&-		&coupled to dust &10$^3$\,km\,s$^{-1}$\\
			& $v_{\rm min}$ 		& - 				&-				&coupled to dust &10$^3$\,km\,s$^{-1}$\\ 
			& $\beta$ 			& - 				&-					&$U(1.5,3.5)$ &\\ 
			& ${F_{\rm 6300}}/{F_{\rm 6364}}$&-				&-			&$U(2.0,3.3)$ &\\ \hline
			\multicolumn{2}{l}{Total number of variable parameters} & 5 & 5 & 10 & \\ \hline
		\end{tabular*}	
	\end{table*}
} \quad

The observational uncertainty in each frequency bin ($\sigma_{\rm obs,i}$) is usually determined when data are reduced and is often included in addition to fluxes in flux-calibrated spectral data files.  However, in a number of cases, particularly in cases of older, archival data, accurate uncertainties are not available.  In these cases, a region of flat continuum may be selected and the observational uncertainty estimated from the variance of fluxes in that region.  A number of different `flat' regions of the spectrum should be sampled and the mean variance calculated.  This value may be used as an approximation to $\sigma_{\rm obs}^2$ which is assumed to be constant over the whole line profile.  Whilst this is an approximation, over the small wavelength ranges of interest for a single line profile, it is generally reasonable to assume that there is little variation in the uncertainty although care should be taken if there was significant contamination to the profile by, for example, sky lines.  Where accurate errors are available, or a full set of raw observations is available such that accurate uncertainties can be calculated, these should be adopted.   The observational error should ideally include accurately calculated uncertainties from as many sources of observational uncertainty as possible (instrumental noise, calibration errors etc.) but particular care should be paid when handling  continuum subtraction.  This can be a significant factor that influences the results of line profile fitting and ideally should be included as a free parameter in any modelling.  This is discussed further in Section \ref{sscn_87A_results}.
 
Each modelled line profile is also inherently uncertain due to the stochastic nature of Monte Carlo simulations.  The Monte Carlo uncertainty can be quantified analytically.  {\sc damocles} propagates weighted energy packets through a dusty medium.  Once it has escaped, the weighted packet is added to the appropriate frequency bin.  Each frequency bin therefore receives weighted packets at a rate that is determined by the properties of the model.  Statistically, this is described by a compound Poisson distribution (i.e. identically, independently distributed weights arrive at a rate described by a Poisson distribution). In the limit of a large number of packets (as here), the compound Poisson distribution can be approximated by a normal distribution with associated Monte Carlo uncertainty in each frequency bin $\sigma_{\rm mod,i}$ described by 

\begin{equation}
\sigma_{\rm mod,i} = f_{\rm obs}  \frac{\sqrt{\sum_{\rm j=1}^{n_{\rm i}}  w_{\rm ij}^2}} { \sum_{\rm i,j}  w_{\rm ij}}
\end{equation}

\noindent where $f_{\rm obs}$ is the total integrated flux of the observed line profile, $n_i$ is the number of packets in bin $i$ and $w_{ij}$ is the weight of the $j^{th}$ packet to arrive in bin $i$. The model flux in the $i^{th}$ frequency bin is given by $f_{\rm mod,i} = f_{\rm obs}  \sum_{\rm j=1}^{n_{\rm i}}  w_{\rm ij}/  \sum_{\rm i,j}  w_{\rm ij}$. The fluxes are therefore scaled such that the total integrated flux of the model profile is equal to that of the observed profile. 

Since both the observational and Monte Carlo uncertainties can be assumed to follow normal distributions, the total error in the likelihood function (see Equation \ref{eqn_likelihood}) therefore also follows a normal distribution and is described by 

\begin{equation}
\sigma_i^2 =  \sigma_{\rm obs,i}^2 + \sigma_{\rm mod,i}^2
\end{equation}

\noindent thus fully defining the likelihood function.

\subsection{Computational implementation}
The Python package `emcee' \citep{emcee} was coupled to the Fortran 95 {\sc damocles} code using the F2PY Fortran-to-Python interface generator \citep{f2py}.  Samples in parameter space are generated in Python, passed to {\sc damocles} where the full Monte Carlo radiative transfer calculation is performed, before the model line profile is passed back to Python. The likelihood and prior are calculated and the algorithm progresses accordingly.  {\sc damocles} is parallelised using OpenMP \citep{OpenMP} and models were run on an 88-core machine with Intel Xeon CPU E5-4669 2.20GHz processors using half its capacity. The most complex, 10-dimensional model took approximately 2 weeks to converge ($\sim$20,000 steps). 

\begin{figure*}
	\includegraphics[clip = true, trim = 0 0 0 0, width = \linewidth]{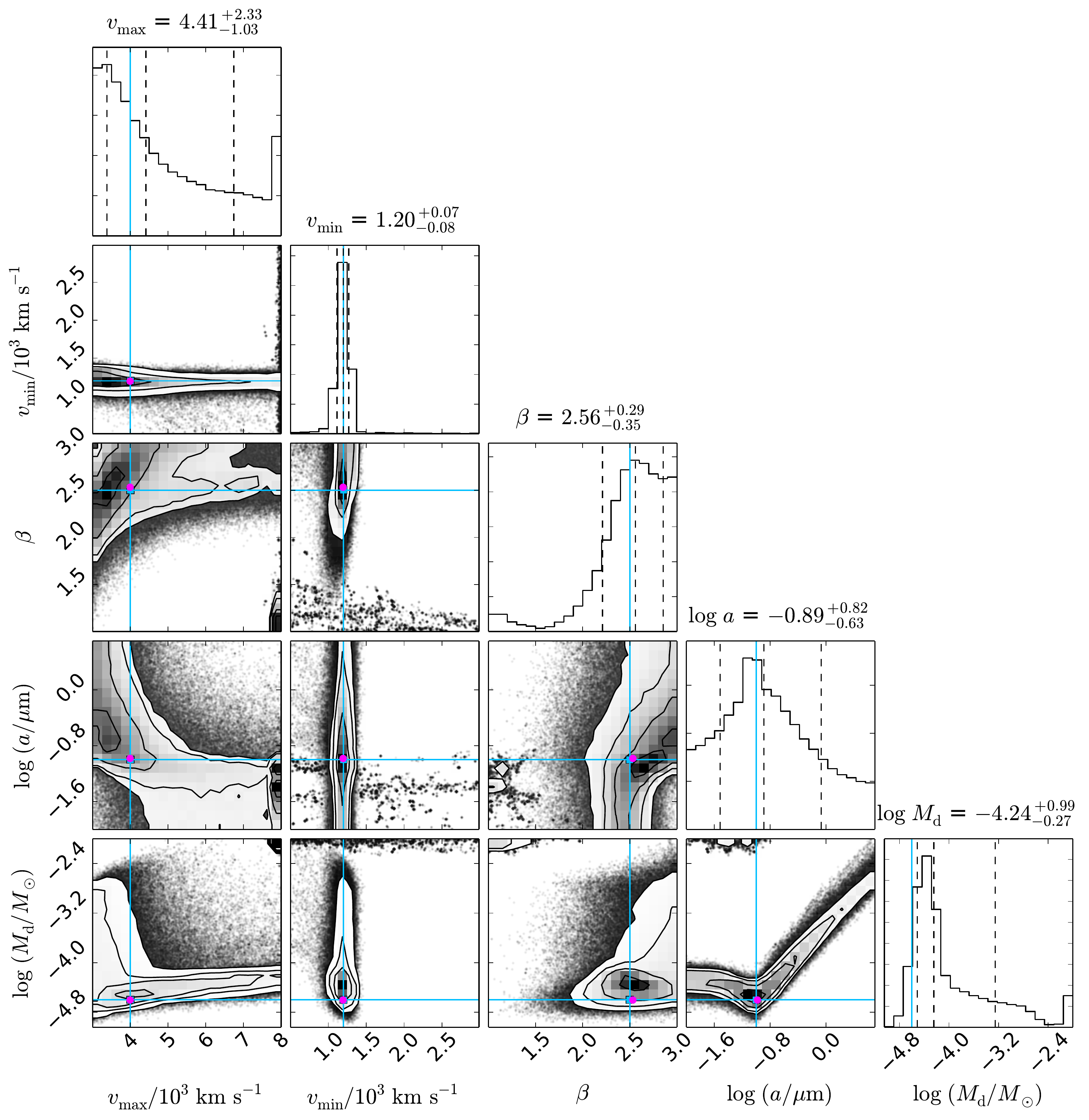}
	\caption{The full posterior probability distribution for the `typical' simulated line profile (model A1).  The known, `true' values used to generate the line profile are marked by the blue cross-hairs and the best-fitting parameter set from the MCMC run is marked with a magenta circle. The adopted priors for this model are presented in Table \ref{tb_synthetic_profiles}. The contours of the 2D distributions represent $0.5\sigma$, $1.0 \sigma$, $1.5 \sigma$ and $2.0 \sigma$,  and the dashed, black vertical lines represent (left to right) the 16$^{\rm th}$, 50$^{\rm th}$ and 84$^{\rm th}$ quantiles of the 1D marginalised probability distributions. }
	\label{fig_run1}
\end{figure*}

\begin{figure*}
	\includegraphics[clip = true, trim = 0 0 0 0, width = \linewidth]{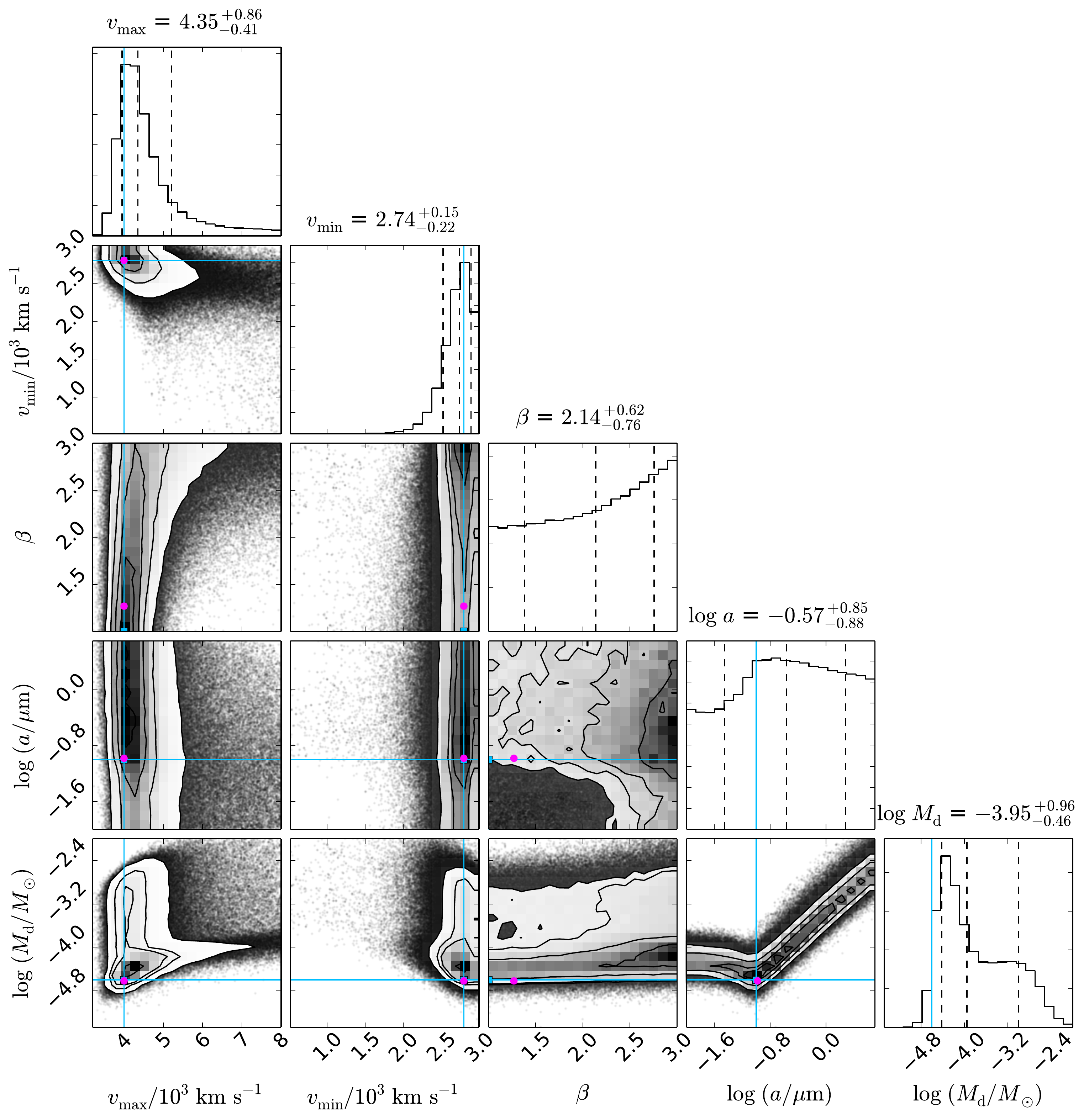}
	\caption{The full posterior probability distribution for the `double peaked' simulated line profile (model A2).  The known, `true' values used to generate the line profile are marked by the blue cross-hairs and the best-fitting parameter set from the MCMC run is marked with a magenta circle. The adopted priors for this model are presented in Table \ref{tb_synthetic_profiles}. The contours of the 2D distributions represent $0.5\sigma$, $1.0 \sigma$, $1.5 \sigma$ and $2.0 \sigma$,  and the dashed, black vertical lines represent (left to right) the 16$^{\rm th}$, 50$^{\rm th}$ and 84$^{\rm th}$ quantiles of the 1D marginalised probability distributions. }
	\label{fig_run2}
\end{figure*}

\begin{figure*}
	\includegraphics[clip = true, trim = 0 0 0 0, width = \linewidth]{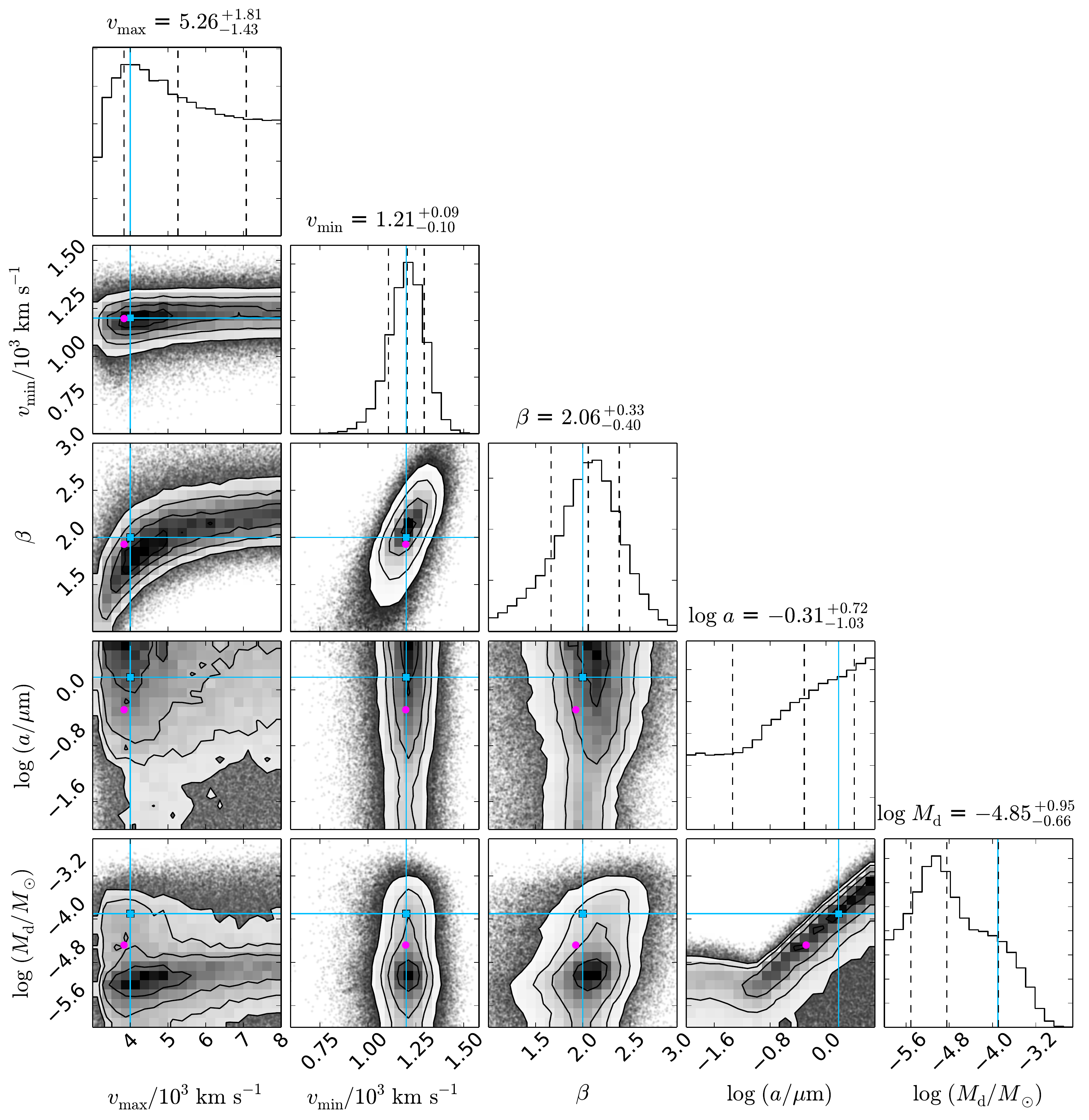}
	\caption{The full posterior probability distribution for the `large grains' simulated line profile (model A3).  The known, `true' values used to generate the line profile are marked by the blue cross-hairs and the best-fitting parameter set from the MCMC run is marked with a magenta circle. The adopted priors for this model are presented in Table \ref{tb_synthetic_profiles}. The contours of the 2D distributions represent $0.5\sigma$, $1.0 \sigma$, $1.5 \sigma$ and $2.0 \sigma$, and the dashed, black vertical lines represent (left to right) the 16$^{\rm th}$, 50$^{\rm th}$ and 84$^{\rm th}$ quantiles of the 1D marginalised probability distributions. }
	\label{fig_run3}
\end{figure*}

\begin{figure*}
	\includegraphics[clip = true, trim = 0 0 0 0, width = \linewidth]{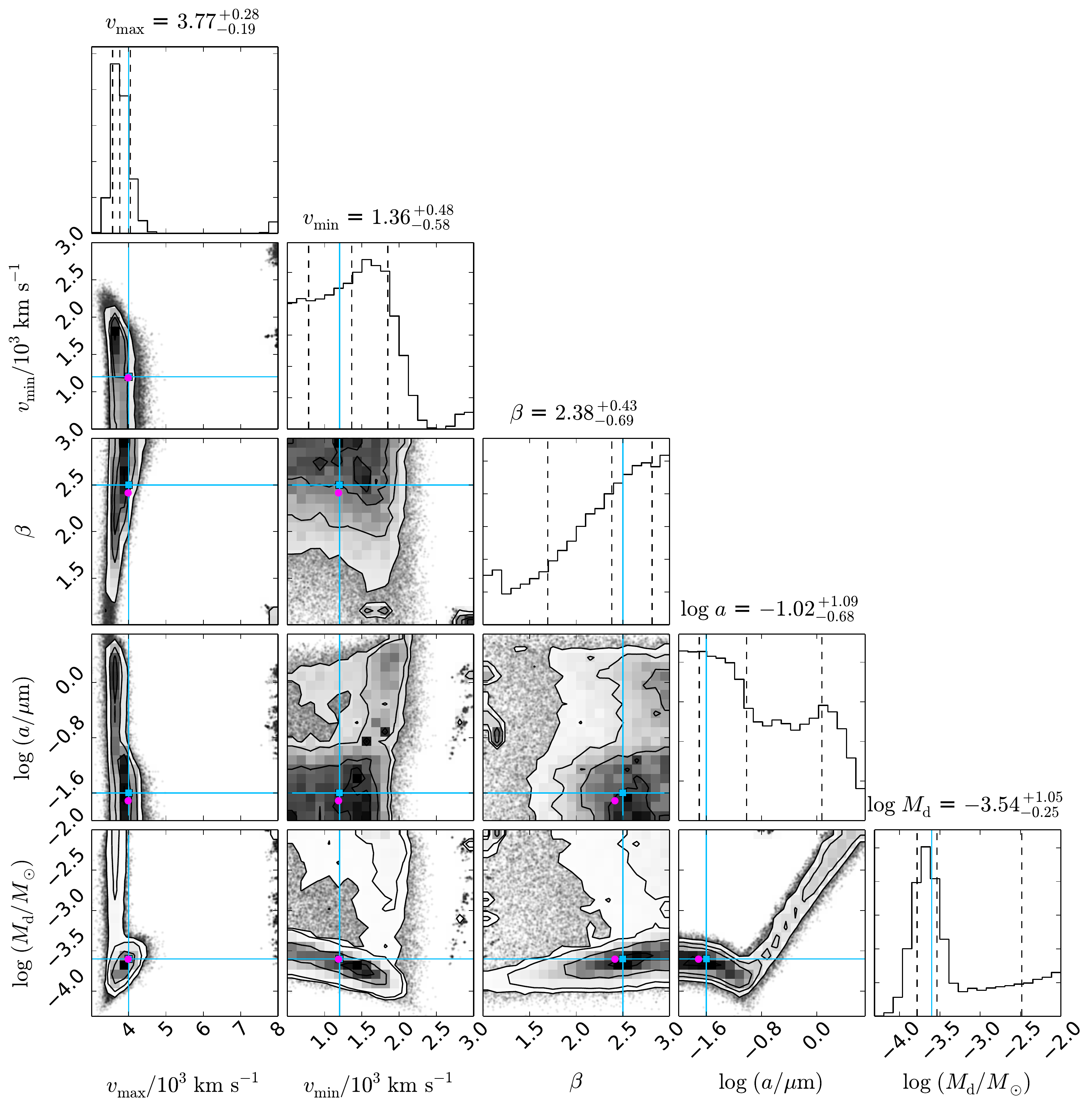}
	\caption{The full posterior probability distribution for the `strongly blueshifted' simulated line profile (model A4).  The known, `true' values used to generate the line profile are marked by the blue cross-hairs and the best-fitting parameter set from the MCMC run is marked with a magenta circle. The adopted priors for this model are presented in Table \ref{tb_synthetic_profiles}. The contours of the 2D distributions represent $0.5\sigma$, $1.0 \sigma$, $1.5 \sigma$  and $2.0 \sigma$, and the dashed, black vertical lines represent (left to right) the 16$^{\rm th}$, 50$^{\rm th}$ and 84$^{\rm th}$ quantiles of the 1D marginalised probability distributions. }
	\label{fig_run4}
\end{figure*}

\section{Results}
\label{scn_results}

\subsection{MCMC models of simulated line profiles (A)}
\label{sscn_results_theoretical}

I initially considered a number of simulated line profiles for which the true parameters were known.  Four simulated line profiles were produced that are similar to the types of asymmetric dust-affected optical and NIR line profiles observed in the spectra of late-time CCSNe with regard to the extent of their asymmetries, their shape and notable features.  The parameters used to generate these line profiles are described in Table \ref{tb_synthetic_profiles} with graphical representations of the geometrical structures presented in Figure \ref{fig_shells} and the profiles presented in Figure \ref{fig_sim_line_profiles}. All four profiles exhibit a blueshifted peak flux due to increased absorption by dust of redshifted radiation and an extended red scattering wing caused by repeated dust scattering events. Three of the profiles also display a `shoulder' or second peak at the position of the minimum radial velocity on the red side. This has been previously noted by B16 and occurs in scenarios with steeper dust and gas density distributions as a result of significant absorption in the central regions of the profile. 

A monochromatic line at 6563\,\AA\ (H$\alpha$) was modelled in each case assuming a post-explosion date of 1000\,d and a symmetric shell ejecta in homologous expansion ($v \propto r$) with maximum velocity at $R_{\rm out}$ of $v_{\rm max} = 4000$\,km~s$^{-1}$.  The models adopted an intrinsic smooth power-law emissivity distribution which was coupled to the square of the density distribution of the dust ($\rho_{\rm d}$) such that for $\rho_{\rm d} \propto r^{-\beta}$ the emissivity distribution followed $i \propto r^{-2\beta}$, as appropriate for recombination lines assuming a constant dust-to-gas mass ratio.  Two schematics that illustrate the structure of the shell geometries and the smooth, radial power-law dust density distributions are presented in Figure \ref{fig_shells} using models A1 and A2 as examples. 

 100\% amorphous carbon grains were used and the optical constants presented by \citet{Zubko1996} were adopted. The physical extent of the ejecta in each model was determined within the code based on the post-explosion time and the specified maximum velocity ($R_{\rm out} = v_{\rm max}t$). The four simulated line profiles that were selected for investigation are presented in Figure \ref{fig_sim_line_profiles} with the parameters used to generate them detailed in Table \ref{tb_synthetic_profiles}. 
 
The ensemble sampler was applied to each of these four simulated profiles.  Five variable parameters were investigated in each case, namely: the maximum velocity ($v_{\rm max}$), the minimum velocity ($v_{\rm min}$), the index of the power-law dust density distribution ($\beta$), the grain radius ($a$) and the total dust mass ($M_{\rm d}$).  These parameters were selected on the basis of previous models (B16, \citet{Bevan2017}) which suggested that they are the  parameters to which a simulated line profile is most sensitive in a spherically symmetric scenario. Whilst dust optical depth and albedo could be substituted for dust mass and grain radius, the latter parameters were chosen in order to allow for more straightforward comparison to other works which present results in these terms. Prior distributions were adopted on all parameters and are described in detail in Table \ref{tb_priors}. Uniform priors were adopted for the maximum and minimum velocities and for the index of the density distribution.  Uniform priors were appropriate for these parameters since I sought to assume minimal prior knowledge and the range of feasible values that these parameters could take was easily encompassed within an order of magnitude.  This was not the case for the dust mass and the grain radius however, both of which could take values within a range covering more than three orders of magnitude.  As a result, these parameters were investigated in log space and uniform priors were adopted for $\log M_{\rm d}$ and  $\log a$. The range of the prior for each parameter was either physically motivated (e.g. the minimum velocity of the expanding ejecta cannot be negative) or was based on realistic values given the observed line profile (e.g. there is no flux detected redwards of 8000~km~s$^{-1}$). The adopted priors were the same for each of the four simulated lines.

In each case, 100 walkers were used and the code was run to convergence, which was determined based on the autocorrelation time. In general, several thousand steps were required to approach convergence.  In all cases, the runs were allowed to continue for several autocorrelation times past this point. An acceptance fraction in the range $[0.2,0.5]$ was required in all cases and in most cases the acceptance fraction was $\sim 0.3$.  

Figures \ref{fig_run1} to \ref{fig_run4} illustrate the results of these models.  For each simulated line profile, a two-dimensional contour plot of the posterior probability distribution for each pairing of the variable parameters is presented. The contours on these plots represent $0.5\sigma$, $1.0 \sigma$, $1.5 \sigma$ and $2.0 \sigma$. Additionally, one-dimensional histograms of the probability density distribution for a single parameter (marginalised over the other parameters) are also presented with the 16$^{\rm th}$, 50$^{\rm th}$ and 84$^{\rm th}$ quantiles indicated, encompassing the central two-thirds of the data. The known, `true' values that were used to generate the simulated line profiles are marked on these plots.  For the sake of comparison, the single best-fitting model was tracked throughout the sampling process and this is also marked on the plots. As expected, the best-fitting model line profiles were virtually identical to the simulated line profiles input into the simulation at the start and so are not presented here.

In most instances, the parameters can be  tightly constrained.  However, there are certain parameters which exhibit a broad posterior probability distribution indicating that the line profile is largely insensitive to variations in this parameter. Dependencies and correlations between the parameters can be observed for some of the parameters, for example the maximum velocity and the density profile (see Fig. \ref{fig_run3}) or the grain radius and the dust mass (see Fig. \ref{fig_run1}).  For the majority of cases, the true values lie very close to or inside the most likely ($1\sigma$) regions of the contour plots.  Where there are exceptions to this, these can be understood as an insensitivity to  a specific parameter on which another parameter is dependent.  In particular, where the dust grain radius cannot be determined from the line profile, the dust mass is likely also to be ill-constrained.  I discuss the reasons for, and implications of, these results in more detail in Section \ref{scn_discussion}.

\subsection{MCMC model of SN 1987A}
\label{sscn_87A_results}

SN~1987A is an extremely well-studied, nearby CCSN that remains critical to our understanding of the formation and evolution of dust in CCSNe. Spectra of SN~1987A have been taken regularly since its outburst on 23 February 1987 and it is an ideal candidate for line profile modelling with asymmetric optical line profiles exhibited from $\sim$650\,d \citep{Lucy1989}.

\begin{figure}
	\includegraphics[clip = true, trim = 60 190 70 205, width = \linewidth]{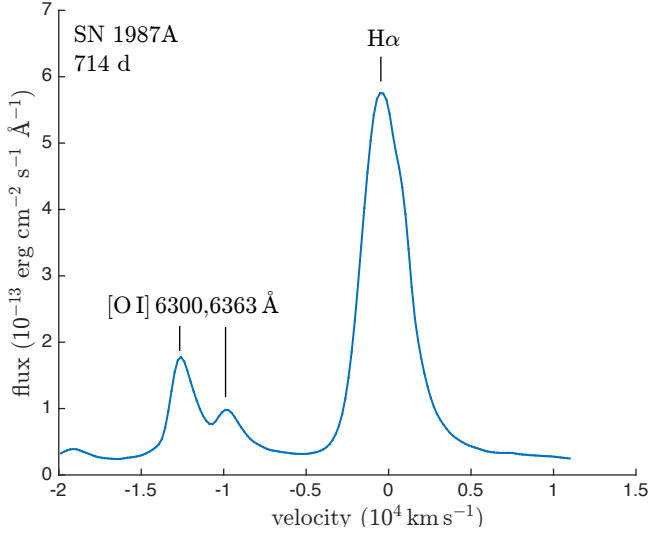}
	\caption{A region of the optical spectrum of SN 1987A at 714\,d post-explosion encompassing the H$\alpha$ line and the [O~{\sc i}]\,6300,\,6363\,\AA\ doublet obtained with the CTIO-1.5m telescope in 1989. The spectrum is centered on zero-velocity at $\lambda = 6563$\,\AA. }
	\label{fig_87A_spectrum}
\end{figure}

\begin{figure*}
	\includegraphics[clip = true, trim = 0 0 0 0, width = \linewidth]{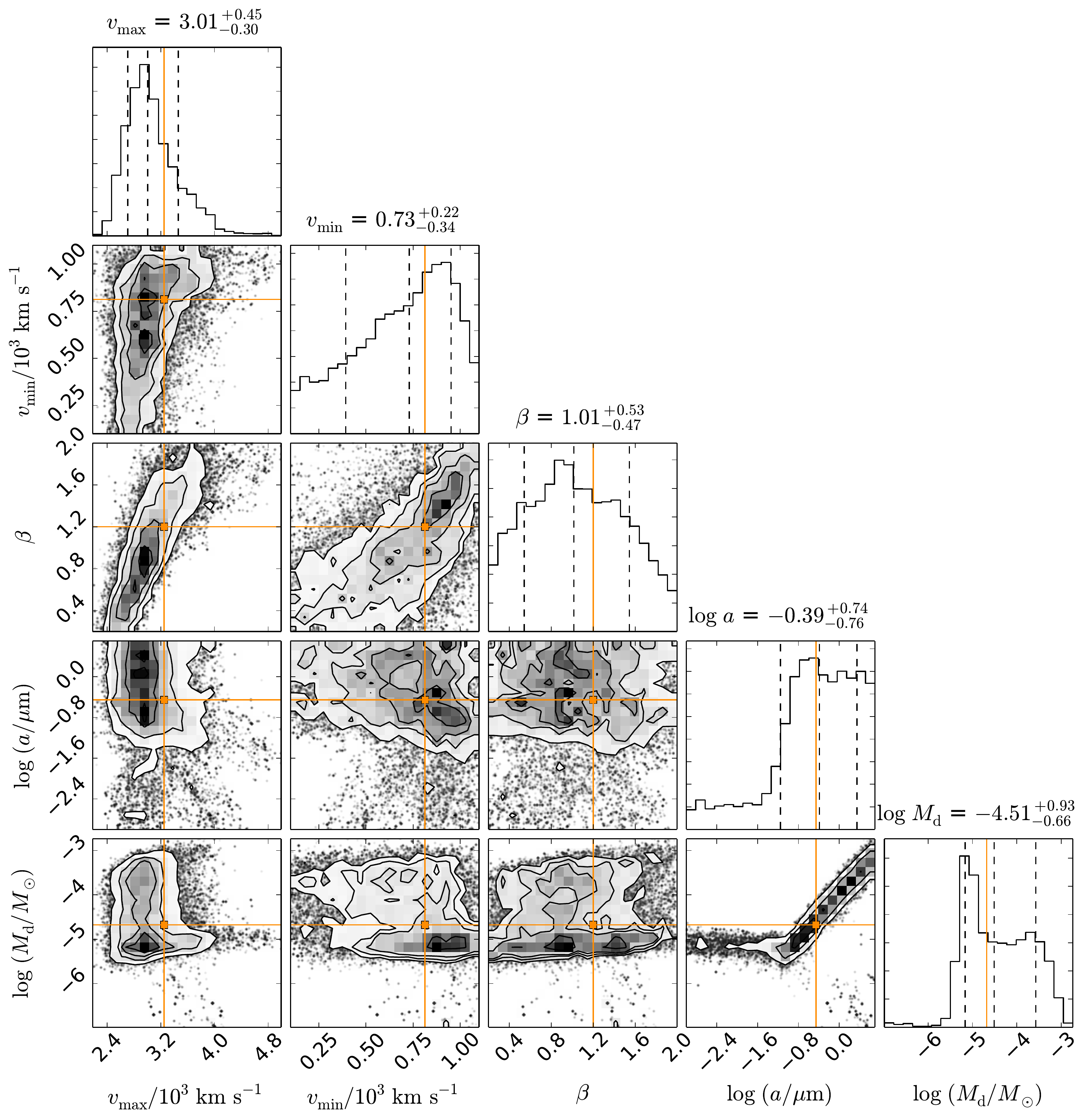}
	\caption{The full posterior probabilty distribution for the smooth, 5-dimensional model of the SN~1987A H$\alpha$ line at 714\,d as described in Section \ref{sscn_87A_results} (model B).  The adopted priors for this model are presented in Table \ref{tb_synthetic_profiles}. The estimated best-fit values from the manual fitting of B16 are marked by the orange cross-hairs. The contours of the 2D distributions represent $0.5\sigma$, $1.0 \sigma$, $1.5 \sigma$ and $2.0 \sigma$ and the dashed, black vertical lines represent the 16$^{\rm th}$, 50$^{\rm th}$ and 84$^{\rm th}$ quantiles for the 1D marginalised probability distributions. }
	\label{fig_87a_smooth}
\end{figure*}

I applied the ensemble sampler to the H$\alpha$ and [O~{\sc i}]\,6300,\,6363\,\AA\ lines of SN~1987A at 714\,d post-outburst.  A region of the optical spectrum that was obtained with the CTIO-1.5m telescope on 6$^{\rm th}$ February 1989 and includes these lines is presented in Figure \ref{fig_87A_spectrum} \citep{Phillips1990}.  The spectrum is available on the CTIO archives. This epoch was selected to revisit due to the high signal-to-noise ratio of the spectrum and the distinct separation of the two features, which is not as clear at later epochs.  Additionally, the relative lack of  contamination of these broad lines by narrow nebular emission makes this epoch particularly attractive.

Both the H$\alpha$ and [O~{\sc i}]\,6300,\,6363\,\AA\ lines have been previously investigated by B16 using {\sc damocles}, who used a systematic, manual approach to determine a best-fitting set of parameters for both clumped and smooth dust density distributions.  I sought to compare the best-fitting parameters that they inferred with the results generated by the automated ensemble sampler.  Additionally, I was interested to understand whether a more sophisticated model that investigates a significantly higher-dimensional variable parameter space could be explored by employing the ensemble sampler. A grid-based or manual approach would not be feasible for higher-dimensional models and this was a primary consideration in the implementation of a MCMC procedure.  I have therefore investigated two models for SN 1987A at 714\,d post-outburst.  

\begin{figure*}
	\includegraphics[clip = true, trim = 0 0 0 0, width = \linewidth]{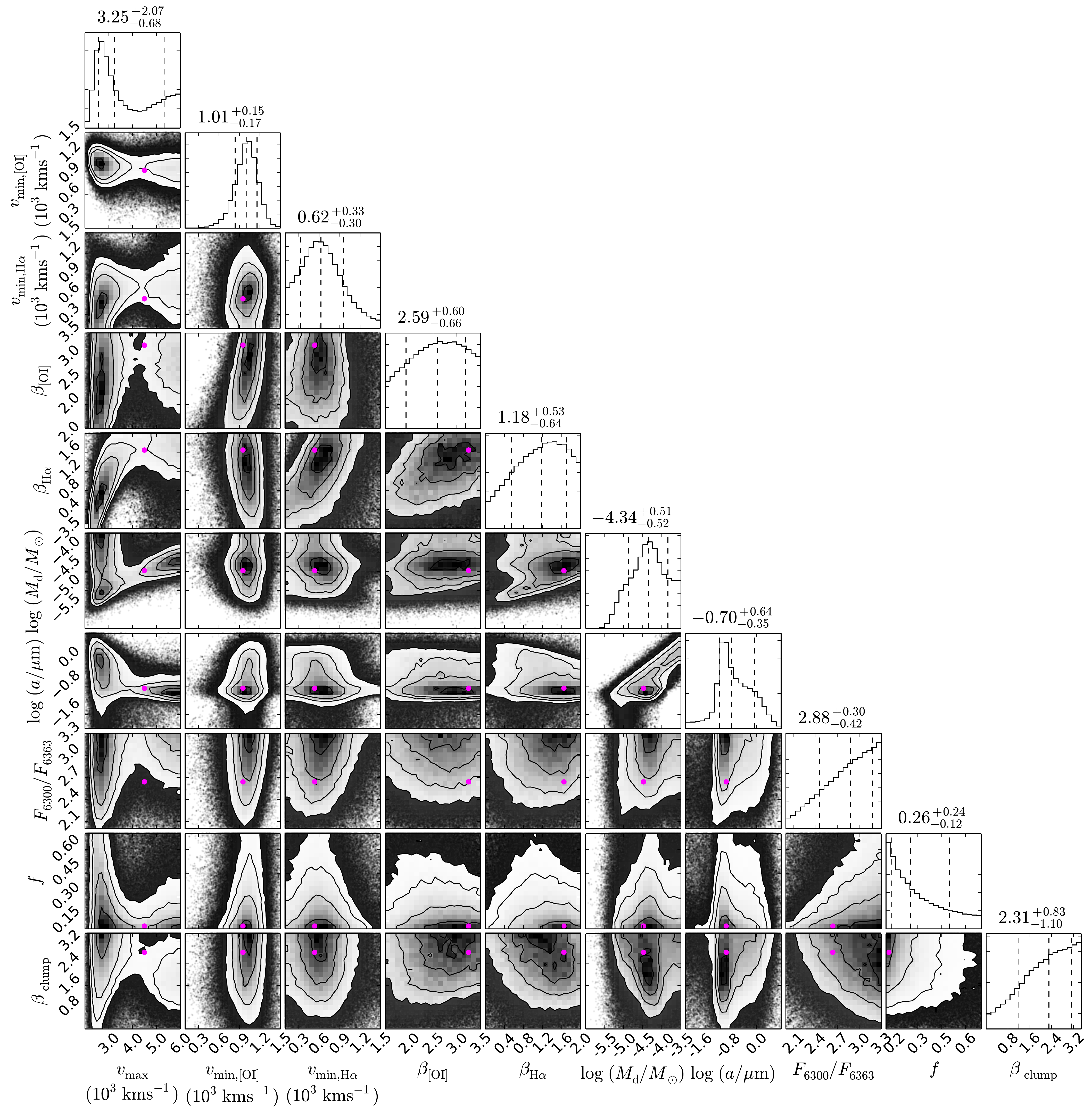}
	\caption{The full posterior probabilty distribution for the clumped, 10-dimensional model of the H$\alpha$ line and [O~{\sc i}]\,6300,\,6363\,\AA\ doublet of SN~1987A  at 714\,d as described in Section \ref{sscn_87A_results} (model C).  The adopted priors for this model are presented in Table \ref{tb_synthetic_profiles}. The contours of the 2D distributions represent $0.5\sigma$, $1.0 \sigma$, $1.5 \sigma$ and $2.0 \sigma$  and the dashed, black vertical lines represent the 16$^{\rm th}$, 50$^{\rm th}$ and 84$^{\rm th}$ quantiles for the 1D marginalised probability distributions. The best-fitting parameter set from the MCMC run is marked with a magenta circle.}
	\label{fig_87a_10D}
\end{figure*}

The first is a 5-dimensional, smooth model that allows for direct comparison with previous results.  The H$\alpha$ line is modelled with a spherically-symmetric, smooth shell distribution.  The power-law emissivity distribution is coupled to the dust density distribution as described in Section \ref{sscn_results_theoretical} and the same dust properties were used (i.e. 100\% amorphous carbon dust with optical constants from \citet{Zubko1996}). This scenario is the same as that adopted by B16 for their smooth model of the H$\alpha$ line at 714\,d.  A five-dimensional parameter space is explored. Uniform priors were adopted for the maximum velocity ($v_{\rm max}$), the minimum velocity ($v_{\rm min}$) and the index of the power-law dust density distribution ($\beta$), and log-uniform priors were adopted for the grain radius ($a$) and the total dust mass ($M_{\rm d}$).  Full details of the priors can be found in Table \ref{tb_priors}. The range of the priors  was kept as wide as possible in an effort to identify any additional maxima in the posterior distribution and therefore to obtain all possible solutions.  In all cases, the best-fitting parameter set identified by B16 lies within the prior range adopted here. 

The modelled profiles were convolved to the resolution of the spectrum (16\,\AA) before the likelihood was calculate and the region between 440~km~s$^{-1}$ and 1400~km~s$^{-1}$ was excluded from this calculation since it is contaminated by the unresolved, narrow, nebular [N~{\sc ii}]\,6583\AA\ emission.   The high signal-to-noise ratio of the spectrum resulted in a negligible observational error as determined by assessing the variance in a flat region of the spectrum. The height of the continuum is a potentially important factor in determining a number of the model properties.  However, a preliminary investigation that included the continuum height as a free parameter revealed an insensitivity to the continuum height and so it was fixed at $2.1 \times 10^{-14}$~erg~cm$^{-2}$~s$^{-1}$~$\AA^{-1}$.

The results of this model are presented in Figure \ref{fig_87a_smooth} with the best-fitting parameters as identified by B16 marked on the probability distributions for comparison.  In all cases, the previous results lie within $1\sigma$ of the marginalised 1D probability distribution and within the 1.5$\sigma$ contour of the 2D joint-probability distributions.  This suggests good agreement between the two approaches but, as can be seen, significantly more information is yielded from the full investigation.  For example, the results indicate that the steepness of the density distribution does not significantly affect the likelihood.  They also highlight the relative insensitivity of the dust mass to all parameters except the grain radius.  The predictably strong correlation between grain size and dust mass, as noted by B16, is clear. However, whilst the grain radius has a fairly well-constrained minimum at around 0.05\,$\mu$m, it is not tightly constrained at larger grain sizes and, as such, constraining the dust mass is difficult without further information.

The second model for SN~1987A treats both the H$\alpha$ and [O~{\sc i}]\,6300,\,6363\,\AA\ lines simultaneously for the first time.  A spherically-symmetric shell-based geometry is once again adopted but, in this more complex scenario, the dust is located entirely in clumps that are stochastically distributed throughout the shell according to a power-law distribution.  The clumps all have equal volume equivalent to a single, cubical grid cell in the simulation of width $R_{\rm out}/25$, roughly consistent with what might be expected from Rayleigh-Taylor instabilities in the ejecta. The total volume of the ejecta occupied by dust clumps is described by the filling factor which is varied between 0.1 and 0.7.  All species (dust, H$\alpha$ and [O~{\sc i}]) extend to the same maximum velocity, but the minimum velocities for H$\alpha$ and [O~{\sc i}]\,6300,\,6363\,\AA\ are separate, variable parameters.  The minimum dust velocity is coupled to that of [O~{\sc i}] since it seems likely that most dust formation is occurring in regions of high metallicity where the constituent ingredients of dust grains are available to condense. All three species follow separate power-law density distributions (with the emissivity following the square of the density as described in Section \ref{sscn_results_theoretical}). Finally, the flux ratio between the 6300\AA\ and 6363\AA\ components of the [O~{\sc i}] doublet is also left as a free parameter.  Intrinsically, the flux ratio is fixed. However, whilst it is assumed that the gas is optically thin, it is possible that there remain some gas optical depth effects that could influence this ratio and it is therefore included for clarity. This yields a total of 10 variable parameters which are summarised, along with the adopted priors for each parameter, in Table \ref{tb_priors}.  

The ranges of the adopted priors were motivated by the previous results of B16 and the results from the previous smooth 5D model.  For certain parameters (the dust mass and grain radius), the range was restricted slightly relative to the 5D simulation, without significant loss of information, in order to speed up the calculation. Physical factors also dictated the adopted ranges, e.g. the flux ratio $F_{\rm 6300}/F_{\rm 6363}$, which was capped at 3.3 since the theoretical value for an optically thin medium is 3.1 \citep{Storey2000}, and the filling factor $f$, which must clearly be in the range $[0,1]$. The likelihood was calculated as per Equation \ref{eqn_likelihood} but, for these purposes, the [O~{\sc i}]\,6300,\,6363\,\AA\ was scaled to the same peak flux as the H$\alpha$ line in order to ensure that both features were weighted equally.  This aside, the adopted procedure for this model was identical to that of model B.  

The results of this 10D simulation are presented in Figure \ref{fig_87a_10D}. A significant quantity of information is contained in this figure but it is of particular interest to note that the majority of parameters have been constrained and follow a distribution with a single peak.  The probability distribution peaks at an extreme of the range in the cases of the flux ratio $F_{\rm 6300}/F_{\rm 6363}$, the filling factor $f$ and the clump number density distribution which is specified by $\beta_{\rm clump}$.  The line profile is not highly sensitive to the density distribution of any species but the minimum and maximum velocities can be restricted to a relatively narrow range, regardless of the values of the other parameters.  Of most interest however, is the strongly-peaked marginalised 1D probability distribution for the grain radius suggesting a large grain radius of the order of $\sim0.2\,\mu$m. This has allowed the dust mass to be similarly constrained with the marginalised probability distribution yielding a $1\sigma$ range spanning only one order of magnitude. The best-fitting parameter set is marked in Figure \ref{fig_87a_10D} for comparison and the corresponding line profile is presented in Figure \ref{fig_10D_bestfit} for the purposes of illustrating the goodness-of-fit.

I discuss these results further in the context of dust formation in SN~1987A and other CCSNe in Section \ref{scn_discussion}.

\section{Discussion}
\label{scn_discussion}

\subsection{Theoretical models}
\label{sscn_theoretical_discussion}

Of primary interest in investigating this approach to modelling asymmetric line profiles is whether the Bayesian methodology offers additional insight or rigour in comparison to manual or grid-based frequentist fitting.

\begin{figure}
	\includegraphics[clip = true, trim = 40 0 60 20, width = \linewidth]{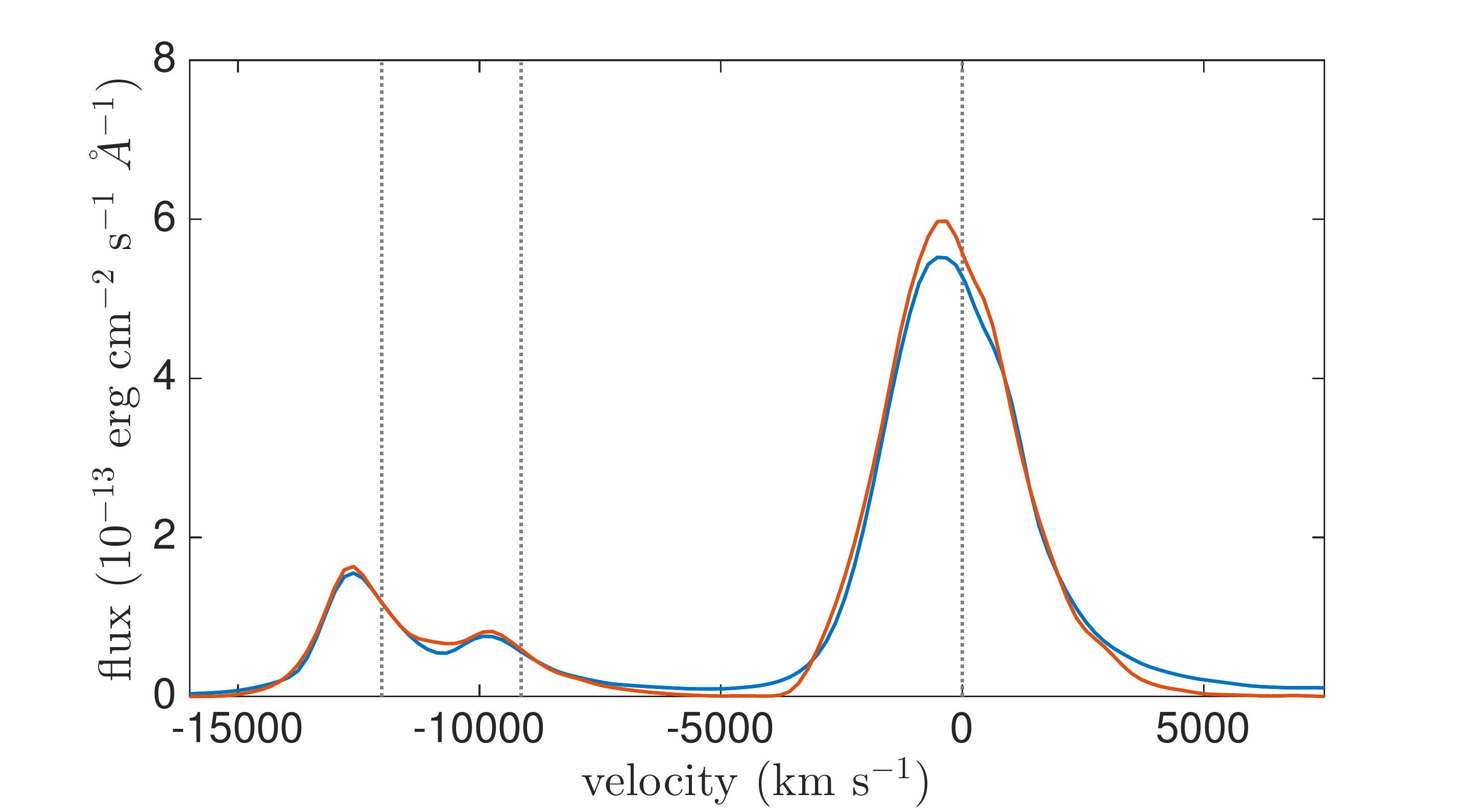}
	\caption{\textit{Blue:} The CTIO spectrum of SN~1987A at 714\,d encompassing the H$\alpha$ line and [O~{\sc i}]\,6300,\,6363\,\AA\ doublet. \textit{Red}: The best-fitting model from the 10-dimensional MCMC run (model C). Vertical, dashed, black lines indicate (left to right) zero-velocity for $\lambda = 6300$\,\AA, 6363\,\AA\ and 6563\,\AA. }
	\label{fig_10D_bestfit}
\end{figure}

\begin{figure}
	\includegraphics[clip = true, trim = 20 0 40 20, width = \linewidth]{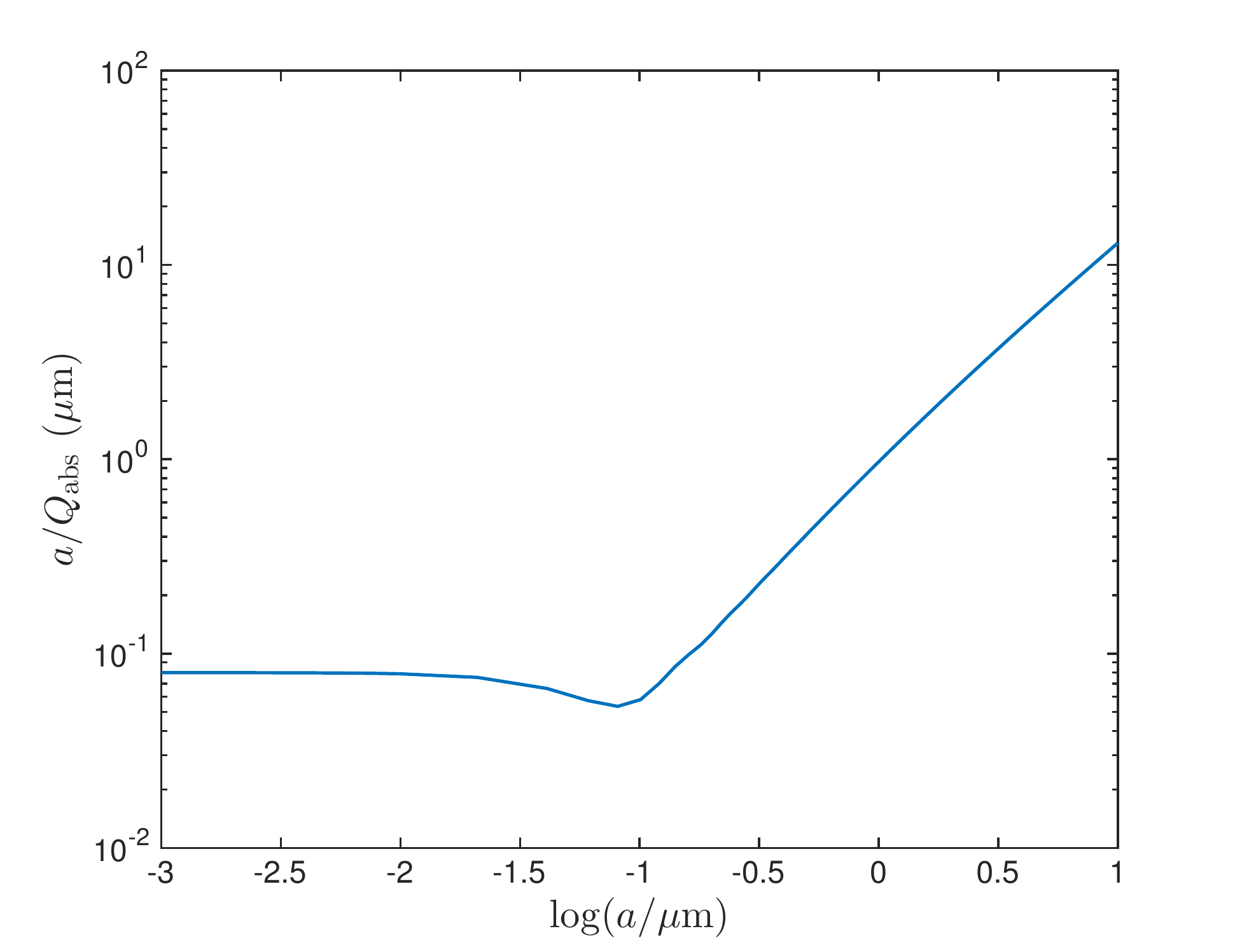}
	\caption{The relationship between the quantity $a/Q_{\rm abs}$ and $a$ at fixed $\lambda = 6563$\,\AA\ as calculated for amorphous carbon using Mie theory, where $a$ is the grain radius and $Q_{\rm abs}$ is the absorption efficiency. $a/Q_{\rm abs}$ is proportional to the dust mass for fixed optical depth, dust density and ejecta size. This relationship is therefore an approximation to the more complex dependency between $a$ and $M_{\rm d}$ seen in the posterior distributions in Figures \ref{fig_run1} to \ref{fig_run4}.}
	\label{fig_a_vs_q}
\end{figure}

The approach was initially tested against four simulated example line profiles that had been generated using {\sc damocles} and that exhibited different shapes and features (models A1~--~A4, see Figure \ref{fig_sim_line_profiles}). The posterior probability distribution for each of these models is presented in Figures \ref{fig_run1} to \ref{fig_run4} with the true parameters used to generate the profiles and the best-fitting parameter set from the chain indicated on the plots. A good test that the Bayesian calculation is being performed correctly is that the best-fitting model from the MCMC chain is, in nearly all cases, in broad agreement with the true values.  The resulting posterior successfully characterises the likelihood of a given parameter over a range of values and exhibits a single-peaked marginalised 1D probability distribution for most parameters, as well as revealing dependencies and correlations between parameters. The known, true values generally lie within the $1\sigma$ contour as would be expected.  Below, I discuss a number of interesting results from the simulated line profile modelling.

\subsubsection{Maximum velocity, $v_{\rm max}$}

The maximum velocity of an emitting species is generally inferred from a line profile as the velocity at which the flux on the blue side of the profile goes to zero.  In the case of models A2 and A4, this approach would yield a reasonable estimate of the maximum velocity in agreement with the posterior distribution, which tightly constrains the value of the maximum velocity in both cases. However, models A1 and A3 (see Figure \ref{fig_run1} and \ref{fig_run3}) illustrate that it is not necessarily straightforward to determine the maximum velocity by eye. The transition between the blue wing of the emission line and the continuum (at zero flux in this theoretical scenario) is smooth for this steep emissivity distribution which makes it difficult to determine the exact velocity at which the transition occurs.  This issue is exacerbated for real observations where noise further obscures the inflection point.  

The full Bayesian calculation illuminates the relative likelihoods of different maximum velocities, considered both in isolation from the other parameters (via the 1D marginalised posterior) and jointly with the other parameters.  It highlights, for example, the postive correlation between the maximum velocity and the emissivity distribution ($i \propto r^{-2\beta}$). Similarly, for model A1 (Figure \ref{fig_run1}), the relationship between grain radius and the maximum velocity is clear.  This can be interpreted as due to the fact that, for amorphous carbon grains, the single-scattering albedo is a monotonically increasing function of grain radius for fixed $\lambda = 6563$\AA. This results in a red scattering wing extending to higher velocities than those observed on the blue side.  This feature of the line profile could be approximated, for small grains, by adopting larger maximum velocities.  This relationship is clearly revealed by the Bayesian calculation whilst the results still prefer the correct grain radius of  $\sim 0.1\mu$m and maximum velocity of approximately  $\sim3250 - 3750$\,km~s$^{-1}$.

Determining the maximum velocity accurately is particularly important since it determines the  size of the ejecta and therefore has a significant effect on the overall dust optical depth to which radiation is exposed for a given dust mass. 
The co-dependence of the maximum velocity with several of the other parameters is handled rigorously by the ensemble sampler and can be easily quantified and communicated via the posterior distribution.

\subsubsection{Minimum velocity, $v_{\rm min}$}

The range of viable values for the minimum velocity can be very narrowly constrained in all optically thin cases (i.e. A1~--~A3).  It is only when the dust becomes significantly optically thick that the minimum velocity becomes harder to determine. In practice, it is normally the case that the blue-shifted peak flux is coincident with the minimum velocity of the emitting ion.  In this case, an asymmetry is observed as a result of absorption in the central regions causing an intrinsically flat-topped, boxy profile (as is produced by an expanding shell) to peak sharply at the blue `corner' of the flat top.  A secondary peak coincident with minimum velocity on the red side is also a possibility \citep[B16,][]{Bevan2017}.  These peaks are important for determining the minimum velocity.  Where dust optical depths are high enough that the peak flux  shifts beyond the minimum velocity (see simulated profile A4, lower right panel of Figure \ref{fig_sim_line_profiles}), there is less information available in the profile to constrain the minimum velocity. The results for A4 (Figure \ref{fig_run4}) suggest that the minimum velocity cannot exceed $\sim$2000\,km\,s$^{-1}$, presumably because the profile would become too wide, but yield similar likelihoods for all $v_{\rm min}<2000$\,km\,s$^{-1}$ with a broad peak centered around $\sim 1600$~km~s$^{-1}$.

\subsubsection{Density distribution index, $\beta$}

The steepness of the density distribution (and hence also the emissivity distribution) is not tightly bound for any of the simulated lines.  However, this is most noticeable in run A2, where there is only a little variation in the likelihood of $\beta$ across the full range explored.  The width of an intrinsic flat-top profile at its peak is determined by the minimum velocity but the shape of the wings is determined (for homologous expansion) by the steepness of the power-law emissivity distribution.  Where this is steeper, the profile appears more concave in its wings.  In A3, only a small fraction of the profile is in the wings, with the majority of the width of the profile arising from the intrinsically flat-topped region. As a result, there is limited information in the profile to allow $\beta$ to be determined. However, the other parameters can still be reasonably estimated by marginalising over $\beta$ since they are not strongly dependent on it.

This is a good illustration of the fact that, under certain conditions, observed line profiles will not contain sufficient information to determine some of parameters of interest.  Even in this case, however, it may be possible to constrain the other parameters by marginalising over these less sensitive parameters. The full posterior distribution clarifies the sensitivity of the line profile to the variable parameters.

\subsubsection{Grain radius, $a$, and dust mass, $M_{\rm d}$}

The strong correlation between grain radius and dust mass is recovered by the Bayesian approach. The absorption and scattering efficiencies of dust grains of any species depend strongly on the grain radius, as does the cross-sectional area available for interaction.  As a result, there is a strong relationship between the opacity and the grain radius, and hence also the required dust mass and the grain radius. 

By making a number of assumptions,  the relationship between dust mass and grain radius can be determined analytically for a simplified version of the scenarios modelled in A1~--~A4. bf We can then compare the 2D marginalised likelihood distributions for dust mass and grain radius to this analytic relationship as a test of the Bayesian approach.
We consider the dust number density to be independent of radius ( which is not the case for models A1~--~A4). The dust optical depth at a given wavelength is then $\tau_{\rm d} = Q_{\rm ext} \pi a^2  n_{\rm d} R$, where $Q_{\rm ext}$ is the extinction efficiency, $n_{\rm d}$ is the dust number density and $R$ is the distance to be traversed by the photon.  
We also have the total dust mass described by $M_{\rm d} = Vn_{\rm d} \frac{4\pi a^3}{3}\rho_{\rm g}$, where $V$ is the volume of the ejecta, $\rho_{\rm g}$ is the mass density of a dust grain and other parameters are as previously defined. 
We require a specific dust optical depth in order to reproduce a line profile.  If we additionally assume that the physical extent of the ejecta is fixed and that the dust is entirely absorbing (such that $Q_{\rm ext} = Q_{\rm abs}$, the absorption efficiency), then we can conclude that $M_{\rm d} \propto a/Q_{\rm abs}$.  
At $\lambda = 6563$\AA, H$\alpha$, we can determine the exact correlation between grain radius $a$ and $a/Q_{\rm abs}$ using Mie theory.  The resulting relationship for amorphous carbon grains (presented in Figure \ref{fig_a_vs_q}) is echoed in the joint 2D posterior distribution of $\log a$ and $\log M_{\rm d}$ in all theoretical runs as expected (A1\,--\,A4; see Figures \ref{fig_run1} to \ref{fig_run4}).

However, the 2D likelihood distributions marginalise over the other parameters and therefore deviate from the analytic relationship derived above. Deviations from this relationship are due to, for example, the polychromatic nature of the transported packets, a dust number density that is non-constant with radius, a significant scattering component to the extinction etc. Silicate grains would be expected to follow a different, more complicated relationship. 

The dependency between grain radius and dust mass has significant implications for determining the ejecta-condensed dust mass via line profile fitting and has been discussed in detail by B16 and \citet{Bevan2017}. By quantifying the posterior distribution, the exact relationship between these two parameters can be understood by marginalising over the other parameters of interest.  If further information can be obtained that would allow the grain radius to be constrained, then the joint probability distribution described by the posterior dictates the required dust mass for a given model.  This grain radius could be estimated from dust emission features or from other techniques such as SED fitting.  The approach used here could also be expanded to include multiple optical or NIR emission lines at a given epoch in order to exploit the wavelength dependence of dust extinction and hence constrain the dust grain radius simultaneously with the other parameters (see Section \ref{sscn_87A_discussion}).

One further implication of the dependency of dust mass on grain radius is that, unless the grain radius can be constrained reasonably tightly, the 1D marginalised dust mass probability distribution will tend towards a specific peak.  This is because there is a wide range of dust grain radii that all yield similar values of $a/Q_{\rm abs}$.  Since the marginalised probability distribution is integrated over the whole prior range of all other parameters, a narrow band of dust masses will naturally be preferred. Care should be taken to ensure that the grain radius has been accurately constrained before inferring the dust mass from the posterior.

It is worth noting that, whilst emphasis is often placed on determining the mass of dust that has formed in the ejecta of CCSNe, and this is therefore naively the most interesting parameter, determining the dust grain radius is critically important in its  own right.  In order to determine how much dust CCSNe can eject into the ISM, we must not only understand how much is formed in the ejecta but also how much is destroyed by shocks, and in particular by the reverse shock that will inevitably pass back through the newly-formed dust \citep{Temim2015,Bocchio2016,Dwek2016}.  The rate of destruction of dust grains by sputtering in shocks is independent of the size of the grain  and, as such, the initial size of the dust grain is critical to understanding whether or not it will survive into the ISM or will eventually be destroyed \citep{Barlow1978}.

\subsection{Application to SN 1987A at 714\,d}
\label{sscn_87A_discussion}

The need to isolate the grain radius motivated the production of two different models of SN~1987A, one significantly more complex that the other.  The initial smooth model in five dimensions (model B) couples the H$\alpha$ emissivity distribution with the dust density distribution and is analogous to the models of H$\alpha$ produced by B16.  Their results are indicated on the posterior distribution which is presented in Figure \ref{fig_87a_smooth}. They are generally in good agreement with the results produced by the ensemble sampler.  However, additional insight is gained into the range of viable values for the maximum and minimum velocities, with the most likely regions of parameter space leaning towards a slightly lower maximum velocity at 3000\,km\,s$^{-1}$ (compared to the B16 estimate of $\sim$3250\,km\,s$^{-1}$ ) and a slightly higher minimum velocity at $\sim$900\,km\,s$^{-1}$ (compared to the B16 estimate of 813\,km\,s$^{-1}$). The steepness of the emissivity distribution is not tightly established but does not affect the ability of the sampler to constrain the other parameters.  

Of most interest, however, is the estimation of the dust grain radius and the dust mass. As previously discussed, there is a strong correlation between these parameters.  Since there is a wide range of small dust grain radii that result in similar dust mass estimates, there is a peak in the marginalised dust mass probability distribution that suggests a dust mass of $\sim$10$^{-6}$\,M$_{\odot}$.  We can infer that this is likely the case if only small dust grains are present in the ejecta.  However, the dust grain radius is not tightly constrained by the smooth fitting, with a wide range of values $>$0.15\,$\mu$m yielding similar probabilities. 

Model C is significantly more detailed and includes additional variable parameters resulting in a 10-dimensional parameter space.  All of the dust is located in clumps. This is a more realistic dust distribution than a smooth radial power-law; dust has been observed to be located in clumpy or filamentary structures in a variety of different CCSNe and remnants \citep{Barlow2010,Gomez2012,Temim2012}.  In addition to a higher-dimensional parameter space, the more complex model also treats both the H$\alpha$ line and the [O~{\sc i}]\,6300,\,6363\,\AA\ doublet simultaneously.  By providing the sampler with more data, and in particular two lines separated in wavelength space, the grain radius can be reasonably constrained.  The median grain radius is $\sim 0.2\,\mu$m, with a maximum at $1\sigma$ of $\sim 1\,\mu$m.  This yields a dust mass that is constrained to within one order of magnitude, with a median dust mass of $\sim 4.5 \times 10^{-5}$\,M$_{\odot}$ and a maximum dust mass at $1\sigma$ of $\sim 1.5 \times 10^{-4}$\,M$_{\odot}$. These estimates are very similar to the separate H$\alpha$ ($5.5 \times 10^{-5}$\,M$_{\odot}$) and [O~{\sc i}]\,6300,\,6363\,\AA\ ($2.0 \times 10^{-4}$\,M$_{\odot}$) estimates by B16 for a grain radius of $0.6\mu$m.  However, they are somewhat lower than the dust mass estimates inferred from radiative transfer models of the SED of SN~1987A presented by \citet{Wesson2015} for this epoch, who deduce a dust mass of $1.0 \times 10^{-3}$\,M$_{\odot}$ at 615\,d post-outburst.  This  discrepancy may be a result of their adoption of an MRN dust grain radius distribution ($n(a) \propto a^{-3.5}$ for $0.005<a<0.25$, \citet{Mathis1977}) or the assumption here of a single grain size.  

This more complex model, which is clearly still a simplification of a highly complicated reality, yields considerable insight into the relative likelihoods of the velocity distributions of the different species and the mass and grain radius of the dust in the ejecta.  Additionally, however, we also gain insight into other properties of the geometry of the nebula at this epoch. The results suggest that the clumps are all likely concentrated towards the central regions (high $\beta_{\rm clump}$) and occupy only a small fraction of the total volume of the ejecta (low $f$).  Similarly, they indicate that the [O~{\sc i}] is also concentrated towards the central regions (median $\beta_{\rm [OI]}$ of 2.59) with the hydrogen more diffusely distributed (median $\beta_{{\rm H}\alpha}$ of 1.18). This suggests a geometry that would be consistent with observations of SN~1987A obtained by \citet{Abellan2017} using the Atacama Large Millimeter Array (ALMA).  These spatially-resolved observations of IR lines of CO and SiO reveal that both species are concentrated in the inner ejecta and occupy a clumpy distribution, suggesting that the heavier elements, and in particular oxygen, are likely located in these central regions.

The results would also be consistent with the structures and geometries predicted by hydrodynamic explosion models of CCSNe \citep{Hammer2010,Wongwathanarat2015}.  These models predict that, at very early times, only a few seconds after the explosion, the heavier elements are  mostly located within the central regions of the ejecta with small clumps of fast-moving material escaping at higher velocities.  A more expansive, more diffuse hydrogen envelope is also present.  Once homologous expansion has set in, the geometry will remain self-similar for many hundreds of years, assuming that there is no encounter with significantly dense circumstellar material, and so it may not be unreasonable to compare these results.

\subsubsection*{}

Supernovae and supernova remnants are highly complex objects.  I have not included in my models different dust species, nor dust grain size distributions, and I have also restricted my investigations to geometries that are, with the exception of a stochastically generated dust clump distribution in one case, spherically symmetric.  These are important factors that should be explored in future work.  Similarly, I have explored only a few particular models.  The results of these analyses do not make comment on the validity of the model itself, rather on the relative likelihoods of the parameters given that particular model.  Care should be taken in the future to assess the applicability of a given model and whether dust formation represents the most likely explanation for the observed properties of a given line profile.  This had already been established from previous work in this case (B16).  The application of a Bayesian procedure may prove useful in this regard also since it lends itself well to quantified model comparison.  However, the above results illustrate the overall power of this methodology to constrain parameters, identify the parameters to which the line profile is sensitive and characterise dependencies between the parameters.  Most importantly, I am able to investigate and analyse highly complex models for which manual parameter estimation would be extremely difficult.

\section{Conclusions}
\label{conclusions}

I have applied an affine invariant ensemble sampler to the Monte Carlo radiative transfer code {\sc damocles} in order to explore the variable parameter space in a rigorous fashion and apply a Bayesian methodology to the inference of conclusions from the data.  I have utilised the algorithm presented by \citet{Goodman2010} and implemented by \cite{emcee} in order to create a Fortran-Python hybrid code that is capable of fitting dust-affected optical and NIR line profiles from CCSNe at late stages in their evolution in order to construct a posterior probability distribution.

The code was applied to four different simulated line profiles that were generated by {\sc damocles} in order to represent different sorts of dust-affected line profiles that are observed in the spectra of late-time dust-forming CCSNe.  A smoothly distributed, spherically symmetric geometry was adopted and five variable parameters investigated. The posterior distributions are in good agreement with the known, true parameters and suggest that the methodology is accurate and effective for parameter estimation.  The theoretical runs highlight a number of dependencies between specific parameters.  The power of the Bayesian inferential approach in revealing and quantifying these dependencies is beneficial for future research using this methodology, but also illustrates the need for care when using line profile fitting (or indeed any other method) to estimate model parameters from observations. 

I also revisited the H$\alpha$ line and [O~{\sc i}]\,6300,\,6363\,\AA\ doublet of SN~1987A at 714\,d. A simple model  with five variable parameters analogous to the smooth model of H$\alpha$ investigated by B16 was initially adopted.  I also investigated a significantly more complex model in 10-dimensional parameter space that treated both H$\alpha$ and [O~{\sc i}]\,6300,\,6363\,\AA\ simultaneously.  The dust mass and dust grain radius predictions  are in agreement with the previous manual approach but their relative likelihood is now quantified, as is their dependence on other parameters.  The affine invariant ensemble sampler has proved to be an efficient and effective method to investigate and analyse highly complex models for which manual parameter estimation would be extremely difficult. The Bayesian methodology allows for considerably more insight to be gained and communicated than the previous manual approach and there is significant potential for using this approach to determine accurate ejecta dust masses for a large number of CCSNe.

\section*{Acknowledgements}

AB would like to thank Dr Boris Leistedt for his patient guidance and teaching on astrostatistics and Bayesian methodologies, as well as Dr Roger Wesson, Dr Ilse de Looze and Professor Mike Barlow for many discussions and their help in readying this paper for publication. Many thanks also to the anonymous referee for their helpful suggestions. 

This work was supported by European Research Council (ERC) Advanced Grant SNDUST 694520 and is based on publicly available observations from the archives of the CTIO.

\bibliography{SN_dust_Bayesian_method_v2}
\bibliographystyle{mnras}

\appendix

\label{lastpage}

\end{document}